## Electric-field-tunable intervalley excitons and phonon replicas in bilayer $WSe_2$

Mashael M. Altaiary[1§], Erfu Liu[1§], Ching-Tarng Liang[2], Fu-Chen Hsiao[2,3], Jeremiah van Baren[1], Takashi Taniguchi[4], Kenji Watanabe[5], Nathaniel M. Gabor[1,6], Yia-Chung Chang[2*], Chun Hung Lui[1*]

[1] Department of Physics and Astronomy, University of California, Riverside, CA 92521, USA.

[2] Research Center for Applied Sciences, Academia Sinica, Taipei 11529, Taiwan.

[3] Advanced Semiconductor Device and Integration Laboratory, Department of Electrical and Computer Engineering, University of Illinois at Urbana-Champaign, Urbana, Illinois 61801, USA

[4] International Center for Materials Nanoarchitectonics (WPI-MANA), National Institute for Materials Science, 1-1 Namiki Tsukuba, Ibaraki 305-0044, Japan.

[5] Research Center for Functional Materials, National Institute for Materials Science, 1-1 Namiki, Tsukuba 305-0044, Japan.

[6] Canadian Institute for Advanced Research, MaRS Centre West Tower, 661 University Avenue, Toronto, Ontario ON M5G 1M1, Canada.

[§]Equal contribution.

[*]Corresponding authors. Email: joshua.lui@ucr.edu; yiachang@gate.sinica.edu.tw

**Abstract:** We report the direct observation of intervalley exciton between the Q conduction valley and Γ valence valley in bilayer $WSe_2$ by photoluminescence. The QΓ exciton lies at ~18 meV below the QK exciton and dominates the luminescence of bilayer $WSe_2$. By measuring the exciton spectra at gate-tunable electric field, we reveal different interlayer electric dipole moments and Stark shifts between QΓ and QK excitons. Notably, we can use the electric field to switch the energy order and dominant luminescence between QΓ and QK excitons. Both QΓ and QK excitons exhibit pronounced phonon replicas, in which two-phonon replicas outshine the one-phonon replicas due to the existence of (nearly) resonant exciton-phonon scatterings and numerous two-phonon scattering paths. We can simulate the replica spectra by comprehensive theoretical modeling and calculations. The good agreement between theory and experiment for the Stark shifts and phonon replicas strongly supports our assignment of QΓ and QK excitons.

Transitional metal dichalcogenides (TMDs, e.g. $MoS_2$, $WSe_2$) are semiconductors with multiple electronic valleys and remarkable excitonic states [1-9]. They exhibit intravalley direct band gap in the monolayer limit, but transit to intervalley indirect band gap in the bilayer and multilayers [1,2]. While much research has focused on the intravalley excitons in monolayer TMDs, intervalley excitons in bilayers are also interesting. Compared to the monolayer excitons, the bilayer excitons possess longer lifetime, different valley configurations, and extra layer degree of freedom, all of which can be harnessed for novel excitonics and valleytronics [10-23].

Of the many TMDs, bilayer $WSe_2$ is a model system to explore the bilayer intervalley excitons [21-25]. Natural bilayer $WSe_2$ crystals possess the 2H stacking order with inversion and time-reversal symmetry [Fig. 1 (a, b)] [11,12,26]. Its electronic band structure exhibits valleys at the zone center (Γ), two zone corners (K, K'), and six interior locations (Q) in the Brillouin zone [Fig. 1(c)] [27,28]. In monolayer $WSe_2$, the Q conduction valleys reside at ~30 meV above the K conduction valleys, and the Γ valence valley resides at ~600 meV below the K valence valleys [29,30]; the Q and Γ valleys play no significant role in the excitonic emission. By contrast, in bilayer $WSe_2$, the Q valleys lie at ~200 meV below the K conduction valley, and the Γ valley is only ~140 meV below the K valence valley [Fig. 1(d)] [28-31]; their contribution to the excitonic emission increases significantly compared to the monolayer case. In particular, while the electronic band structure suggests that the QK intervalley exciton ($X_{QK}$) is the lowest and dominant exciton in bilayer $WSe_2$, the QΓ intervalley exciton ($X_{Q\Gamma}$) can compete with $X_{QK}$ due to its possible stronger excitonic effect [25].

The competition between $X_{QK}$ and $X_{Q\Gamma}$ becomes even more intriguing under an electric field. Unlike monolayer, bilayer excitons can possess interlayer electric dipole [13,16,21,26,32-37]. By using the different interlayer Stark effect of $X_{QK}$ and $X_{Q\Gamma}$, one may tune their energy separation and control their population and luminescence. Their different valley configurations can also induce different exciton-phonon coupling [38], further enriching the exciton physics. Given such an interesting theoretical prospect, it is intriguing to explore bilayer TMDs. But experimental studies have been challenging due to the generally poor quality of bilayer TMD samples. Particularly, although prior research has shown some theoretical and experimental evidence of QΓ exciton (*e.g.* in quantum-dot emission) [25], direct observation of the QΓ exciton in pristine bilayer TMDs has never been reported.

In this Letter, we directly resolve the QΓ exciton emission in bilayer $WSe_2$ and conduct comparative electric-field-dependent studies between QΓ and QK excitons. The QΓ exciton lies at ~18 meV below the QK exciton, revealing stronger binding of the QΓ exciton than QK exciton. Under an out-of-plane electric field, both QΓ and QK excitons exhibit strong Stark effect, but the QΓ Stark shift is considerably weaker than the QK shift. By tuning the electric field strength between 0 and 0.22 V/nm, we can switch the QΓ and QK exciton to be the lowest exciton at different field strength. In addition, we resolve a series of phonon replicas for both excitons. Notably, the phonon replicas outshine the primary excitonic emission, and the two-phonon replicas are stronger than the one-phonon replicas. We can simulate the replica spectra by comprehensive theoretical calculation. Our results reveal the existence of numerous two-phonon scattering processes and (nearly) resonant exciton-phonon scatterings in bilayer $WSe_2$, which cause unusually strong two-phonon replicas in the system.

In our experiment, we fabricate dual-gate bilayer $WSe_2$ devices encapsulated by hexagonal boron nitride (BN) on $Si/SiO_2$ substrates [Fig. 2(a)]. The BN encapsulation significantly improves the sample quality and allows for the observation of QΓ exciton. Thin graphite flakes are used as the contact and gate electrodes to further enhance the device performance. By applying voltages of opposite signs and appropriate ratio on the top gate ($V_{tg}$) and bottom gate ($V_{bg}$), we can generate a vertical electric field across bilayer $WSe_2$ while keeping the sample charge neutral (See Supplemental Materials for details of device fabrication and experiment [39]). Fig. 2(b) displays the photoluminescence (PL) map at varying voltage difference ($\Delta V = V_{tg} - V_{bg}$) between the top and bottom gates at temperature $T \sim 15$ K. At low $\Delta V$, we observe multiple cross-shape emission features; such field-dependent cross-shape features have never been reported. At large $\Delta V$, several redshifting lines appear; they should be associated with the QK exciton according to prior research [21]. We further perform second-order energy derivative on the PL map to resolve the fine features [Fig. 2(c)]. From the map of the second-order derivative, we extract the energies of different emission features and plot them as a function of electric field in Fig. 2(d).

We first consider the two sets of highest-energy features, which consist of one cross and two redshifting lines in 1.60 – 1.63 eV [Fig 2(b-d)]. They are significantly weaker than the replica features at lower energy. We tentatively assign them as the primary $X_{Q\Gamma}$ and $X_{QK}$ emission assisted by defect scattering. Defect scattering is considered because direct optical recombination of intervalley excitons is forbidden by the momentum conservation [40,41]. To confirm our assignment, we conduct first-principles calculations on the band structure of bilayer $WSe_2$ under different vertical electric fields [Fig. 1(d-e)] [39]. At zero field, each band is doubly degenerate due to the inversion and time-reversal symmetry of 2H-stacked bilayer $WSe_2$ [Fig. 1(d)]. At the finite electric field, the inversion symmetry is broken, and each band is split into two bands with opposite spin and layer polarizations (except at the Γ point) [Fig. 1(e)]. Consequently, the $X_{Q\Gamma}$ and $X_{QK}$ emission each split into two lines and exhibit different Stark shifts. We have calculated the QΓ and QK interband transition energies (with no excitonic effect) at different electric fields via the density functional theory (DFT) [dashed lines in Fig. 2(d)] [39]. We offset the theoretical transition energies to match the experimental range at zero field and focus our comparison only on the Stark shifts. The theoretical Stark shifts agree decently with our data (the small deviation may come from the change of exciton binding under electric field). Our theory thus supports our assignment of QΓ and QK excitons.

The QΓ exciton exhibit two remarkable characteristics. First, $X_{Q\Gamma}$ is ~18 meV lower than $X_{QK}$ and is significantly brighter than $X_{QK}$. This result is surprising because the Γ valence valley is ~140 meV lower than the K valence valley in the electronic band structure [Fig. 1(d)] [29]; one would naturally expect that the QΓ transition has higher energy than the QK transition. However, further examination of the band structure reveals that the Γ valley has a much larger hole effective mass (~$1.48m_0$) than the K valence valley (~$0.43m_0$; $m_0$ is the free electron mass) [30]. Hence, the QΓ exciton binds more strongly than the QK exciton. The larger exciton binding energy causes $X_{Q\Gamma}$ to have lower energy than $X_{QK}$. This result is significant because the lowest excitonic state plays a paramount role in the excitonic dynamics and luminescence.

Second, the $X_{Q\Gamma}$ Stark shift (~70 meV per 1 V/nm field) is 1.7 ~ 2.7 times weaker than the $X_{QK}$ Stark shift (120 ~ 190 meV per 1 V/nm field). This can be qualitatively understood from their different charge density distribution. Fig. 3 shows our calculated spin-dependent charge density along the vertical direction for electron states at the Q, K, and Γ points. We consider an infinitesimal electric field to break the inversion symmetry. The separated states show strong spin polarization – the electric field essentially splits them into spin-up and spin-down states. The K,

Q, and Γ states show different layer polarization. The Q conduction states show medium layer polarization with an electric dipole moment $p_Q$ = 2.02 atomic unit (a.u.) [Fig. 3(a, b)]. In contrast, the K valence states are strongly localized in opposite layers, giving rise to a large dipole moment $p_K$ = 5.80 a.u. [Fig. 3(c, d)]. In the other extreme, the two Γ valence states show symmetric distribution on the two layers, which produces no layer polarization and zero electric dipole moment ($p_\Gamma$ = 0) [Fig. 3(e, f)]. By combining these dipole moments, we estimate that a QΓ (QK) electron-hole pair has a dipole moment of $p_{Q\Gamma} = p_Q = 2.02$ a.u. ($p_{QK} = p_Q + p_K = 7.82$ a.u.). This simple estimation qualitatively accounts for the different Stark shift between $X_{Q\Gamma}$ and $X_{QK}$ (Quantitative deviation can occur due to excitonic effect and mixing with other bands under electric field).

The different Stark shifts between $X_{Q\Gamma}$ and $X_{QK}$ have an interesting consequence – they allow us to use the electric field to switch the energy order and dominant luminescence between $X_{Q\Gamma}$ and $X_{QK}$. When the electric field is weak (*e.g.* E < 0.02 V/nm), $X_{Q\Gamma}$ lies well below $X_{QK}$ and dominates the emission due to its higher population. As the field increases, $X_{QK}$ gradually redshifts to become lower than $X_{Q\Gamma}$, and the exciton population transfers from $X_{Q\Gamma}$ to $X_{QK}$; this brightens $X_{QK}$ and suppresses $X_{Q\Gamma}$. At strong electric field (*e.g.* E > 0.2 V/nm), $X_{QK}$ dominates the luminescence. Such electric-field switching effect of excitonic emission is not found in monolayer TMDs and may find novel applications for bilayer TMDs.

After we address the primary $X_{Q\Gamma}$ and $X_{QK}$ emission, we turn to their replica emission at lower energy. Fig. 2 shows five $X_{Q\Gamma}$ replicas ($X_{Q\Gamma}^1 - X_{Q\Gamma}^5$) with respective redshift energies 13.0, 27.6, 41.7, 45.7, and 57.7 meV from $X_{Q\Gamma}$ and six $X_{QK}$ replicas ($X_{QK}^1 - X_{QK}^6$) with respective redshift energies 28.8, 32.4, 41.9, 46.2, 49.1, and 58.3 meV from $X_{QK}$ (the uncertainty of these energies is from $\pm 1$ to $\pm 2$; see Table S6 [39]). The replicas exhibit identical Stark shifts to the primary emission lines. We tentatively attribute them to excitonic luminescence assisted by phonon emission, as phonon replicas have been reported in monolayer $WSe_2$ [40-44] and bilayer $WSe_2$ [24]. In our results, the replicas are considerably brighter than the primary emission, indicating that phonon-assisted emission is more efficient than defect-assisted emission in our system.

To identify the phonon replicas, we have calculated the phonon band structure of bilayer $WSe_2$ with a rigid-ion model [39]. Afterward, we calculate the one-phonon replicas for $X_{Q\Gamma}$ and $X_{QK}$ by the perturbation theory. In our theory, the initial excitonic state is scattered to a second excitonic state by emitting a phonon, then the second state decays to emit a photon. The second state is not necessarily a bound exciton, but it must be a momentum-direct exciton in order to emit the photon. Momentum conservation restricts the one-phonon scattering processes to occur only through two paths. For $X_{Q\Gamma}$, the electron may be scattered from Q to Γ by emitting a phonon ($X_{Q\Gamma} \rightarrow X_{\Gamma\Gamma}$), or the hole may be scattered from Γ to Q by emitting a phonon ($X_{Q\Gamma} \rightarrow X_{QQ}$). Similarly, for $X_{QK}$, the electron can be scattered from Q to K ($X_{QK} \rightarrow X_{KK}$), or the hole from K to Q ($X_{QK} \rightarrow X_{QQ}$). As the initial $X_{Q\Gamma}$ and $X_{QK}$ excitons can be associated with any of the six Q valleys and two K valleys, the momentum of the emitted phonons will be near one of the six Q points for $X_{Q\Gamma}$ and near one of the Q or M points for $X_{QK}$. The emitted phonon can come from any phonon branch, so the one-phonon replica spectra can exhibit multiple peaks (see Table S3 and Fig. S10 for details [39]). These one-phonon scattering processes are all non-resonant, so the one-phonon replicas are generally weak.

Fig. 4 compares our calculated one-phonon spectra for $X_{Q\Gamma}$ and $X_{QK}$ with the experimental spectra. We phenomenologically broaden the theoretical spectra by 2 meV to match the experimental peak width. The theoretical one-phonon spectra can account for the $X_{Q\Gamma}^{1}$ peak, which is contributed dominantly by the LA and TA acoustic phonons near the Q points. It can also partially explain $X_{Q\Gamma}^{2}$ and $X_{QK}^{1,2}$. But other replicas ($X_{Q\Gamma}^{3-5}$ and $X_{QK}^{3-6}$) exceed the range of single phonon energy (~37 meV) in $WSe_2$. We would need to consider higher-order scattering processes to explain these replicas.

We have conducted comprehensive calculations on the two-phonon replicas of $X_{Q\Gamma}$ and $X_{QK}$ [39]. In our theory, the initial exciton is scattered to a second excitonic state by emitting a phonon, and afterward scattered to a third excitonic state by emitting another phonon, and the third state decays to emit a photon. The second and third states are not necessarily bound excitons, but the third state may be a momentum-direct exciton in order to emit the photon. In contrast to the one-phonon processes with only two scattering paths, the two-phonon processes have numerous scattering paths. In our calculation, we only consider those scattering paths involving a resonant or nearly resonant exciton-phonon scattering process. In bilayer $WSe_2$, the six Q conduction valleys are energy-degenerate, and $X_{Q\Gamma}$ and $X_{QK}$ are close in energy (~18 meV). So, carrier-phonon scattering between different Q valleys, between the Γ and K valence valleys, or within the same valley, is resonant or nearly resonant. Two-phonon processes involving such (nearly) resonant scattering are expected to contribute dominantly to the replica spectra.

In our survey, there are totally 16 two-phonon scattering paths with a (nearly) resonant component for QΓ exciton. They can be separated into four groups: $X_{Q\Gamma} \rightarrow X_{Q'\Gamma} \rightarrow X_{Q'Q'}$ ; $X_{Q\Gamma} \rightarrow X_{Q'\Gamma} \rightarrow X_{\Gamma\Gamma}$; $X_{Q\Gamma} \rightarrow X_{QK} \rightarrow X_{KK}$; $X_{Q\Gamma} \rightarrow X_{QK} \rightarrow X_{QQ}$ (here Q′ denotes any of the six Q valleys). Similarly, there are 14 dominant paths for the QK exciton, separated into four groups: $X_{QK} \rightarrow X_{Q'K} \rightarrow X_{KK}$ ; $X_{QK} \rightarrow X_{Q'K} \rightarrow X_{Q'Q'}$ ; $X_{QK} \rightarrow X_{Q\Gamma} \rightarrow X_{QQ}$ ; $X_{QK} \rightarrow X_{Q\Gamma} \rightarrow X_{\Gamma\Gamma}$ (see the Supplemental Materials for the explanation of these paths [39]). Momentum conservation restricts the emitted phonons to be near the Γ, K, Q, and M points (Tables S3, S5 [39]). The emitted phonons can come from any phonon branch, so the two-phonon replica spectra are rather complicated.

Fig. 4(c-d) display our calculated and broadened two-phonon replica spectra, which include contributions from all the above-mentioned scattering paths and all phonon branches. By summing the one-phonon and two-phonon spectra, our total theoretical spectra match decently the experimental spectra, including the energy and relative intensity of different replica peaks. From the theoretical results, we can attribute $X_{Q\Gamma}^{1}$ to the one-phonon replica of Q-point acoustic phonons, $X_{Q\Gamma}^{2}$ to a superposition of one-phonon and two-phonon replicas, $X_{Q\Gamma}^{2-5}$ and $X_{QK}^{1-6}$ to two-phonon replicas (see Table S6 for detailed assignments of each replica peak [39]). In particular, $X_{QK}^{1-5}$ are dominantly contributed by two-phonon paths involving the $X_{QK} \rightarrow X_{Q\Gamma}$ transition with acoustic phonon emission, because this transition is strongly resonant when the emitted phonon energy is close to the 18-meV difference between $X_{QK}$ and $X_{Q\Gamma}$ (Table S5 [39]).

We remark that the two-phonon replicas are considerably stronger than the one-phonon replicas. This characteristic is counter-intuitive because second-order processes are usually much weaker than first-order processes. There are two reasons for this unusual phenomenon. First, one of the scattering components in the two-phonon processes is (nearly) resonant. This makes each considered two-phonon process as strong as a non-resonant one-phonon process. Second, the number of (nearly) resonant two-phonon scattering paths (≥14) considerably exceeds the number of one-phonon scattering paths (only 2). After summing the contributions of these two-phonon paths, the total two-phonon spectra become considerably stronger than the one-phonon spectra. Such a characteristic is not found in monolayer $WSe_2$ due to its different band structure. It makes bilayer $WSe_2$ a distinctive system to study novel exciton-phonon phenomena.

In sum, we have measured the electric-field-dependent spectra of QΓ and QK exciton and show their different interlayer Stark shifts. Both excitons exhibit unusually strong two-phonon replicas. In the Supplemental Materials, we also show the Zeeman splitting effect of $X_{Q\Gamma}^{2}$ under magnetic field, from which we deduce a g-factor of $9.1 \pm 0.4$ for QΓ exciton [25]. The interesting interplay between excitons, phonons, and electric and magnetic fields, as demonstrated in our research, makes bilayer $WSe_2$ a distinctive excitonic system with potential applications in excitonics and valleytronics.

**Acknowledgment:** We thank Fatemeh Barati and Matthew Wilson for help in the project, Ao Shi for assistance in manuscript preparation, Kin Fai Mak and Jie Shan for discussion, Yong-Tao Cui for supporting E.L., and Harry Tom for equipment support. C.H.L. acknowledges support from the National Science Foundation Division of Materials Research CAREER Award No.1945660. YCC acknowledges support by the Ministry of Science and Technology (Taiwan) under Grant No. MOST 109-2112-M-001-046. N.M.G acknowledges support from the National Science Foundation Division of Materials Research CAREER Award No. 1651247, and from the United States Department of the Navy Historically Black Colleges, Universities, and Minority Serving Institutions (HBCU/MI) Award No. N00014-19-1-2574. K.W. and T.T. acknowledge support from the Elemental Strategy Initiative conducted by the MEXT, Japan and the CREST (JPMJCR15F3), JST.

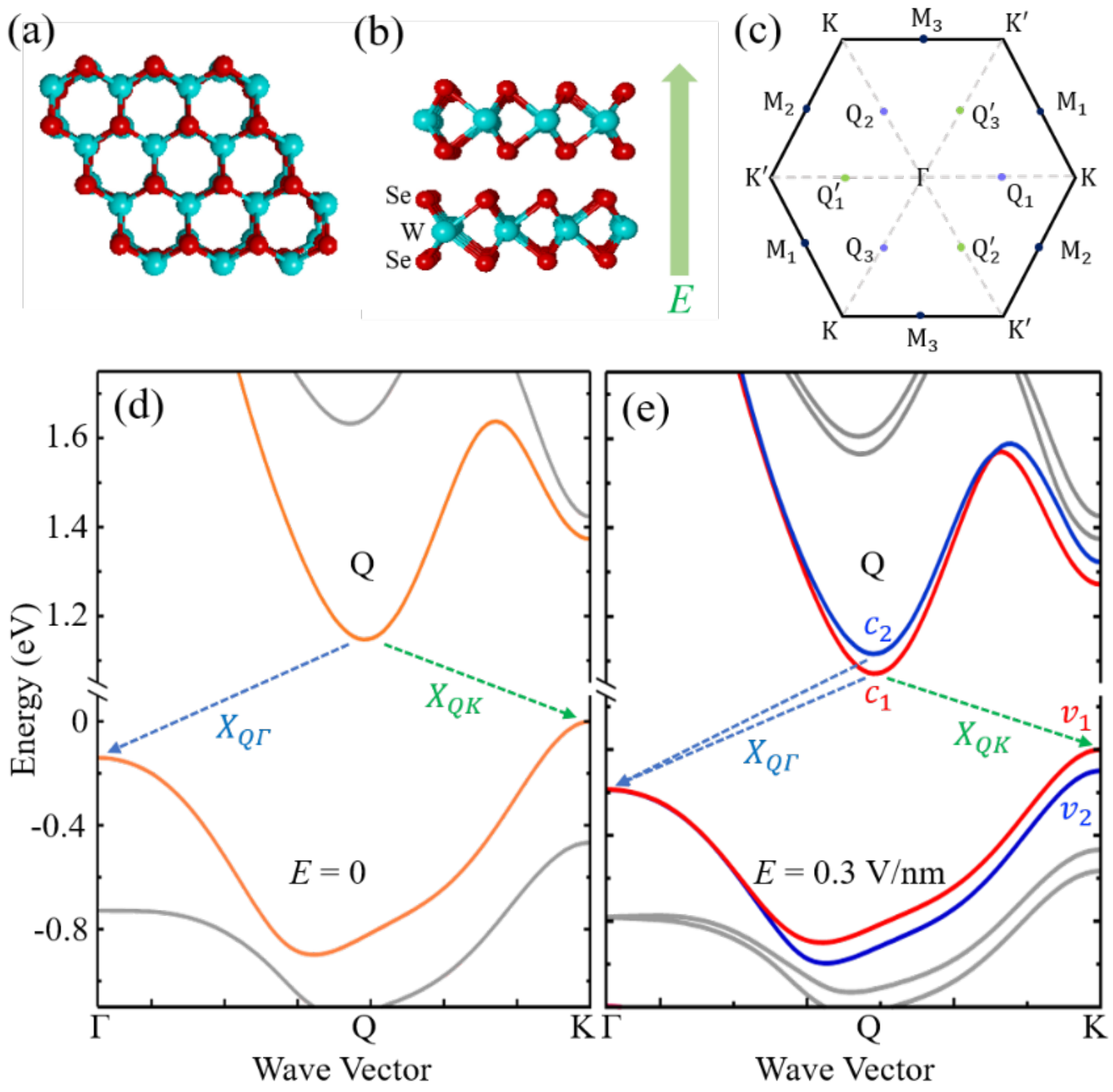

FIG.1. (a) Top view and (b) side view of the crystal structure of 2H-stacked bilayer $WSe_2$. (c) The Brillouin zone of bilayer $WSe_2$. (d-e) Calculated electronic band structure along the K-Γ line at zero (d) and 0.3 V/nm (e) vertical electric field. Each band is doubly degenerate at zero field, but split at finite field (except at the Γ point). We denote the $X_{Q\Gamma}$ and $X_{QK}$ emission and the dominant spin-up (spin-down) polarization by red (blue) color.

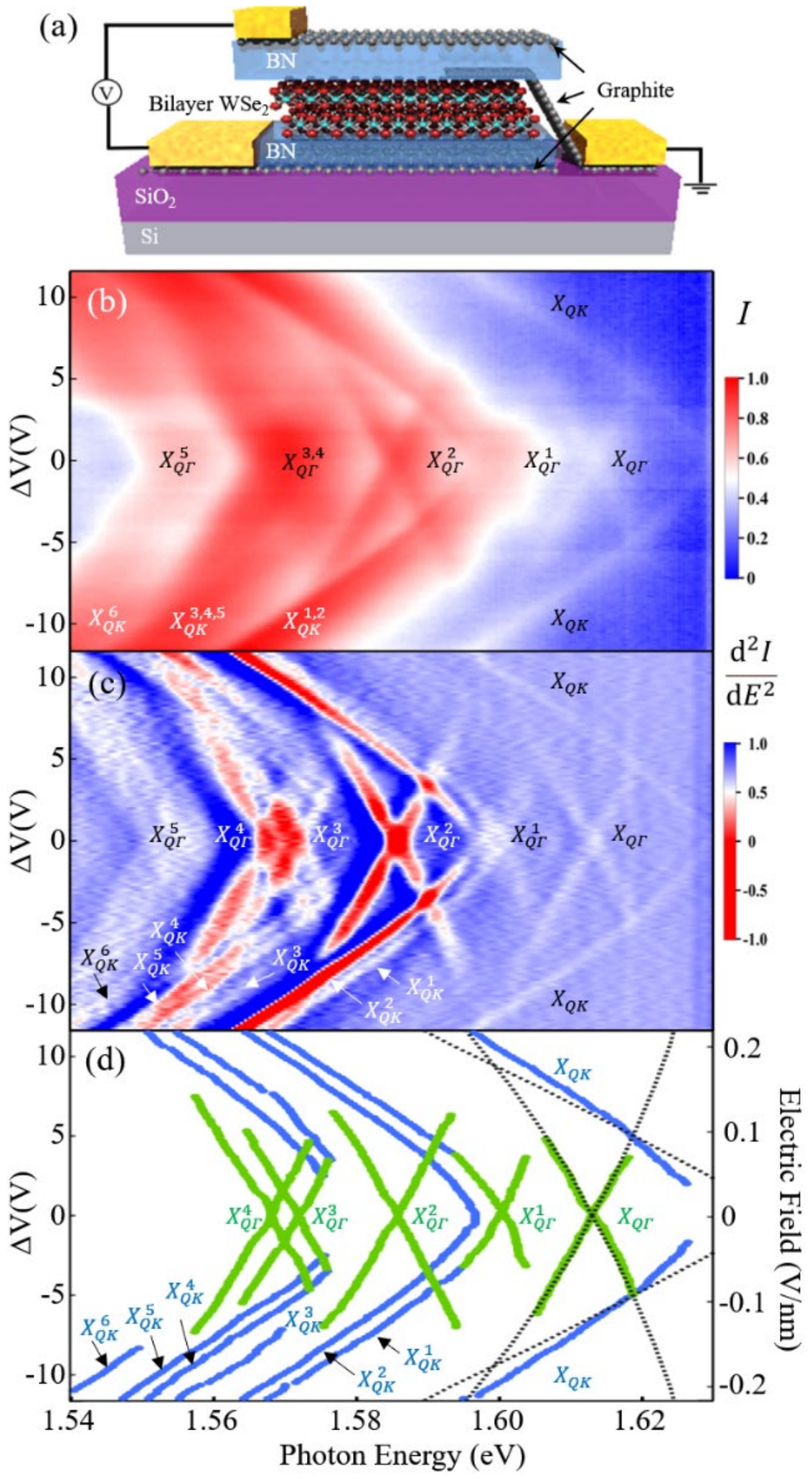

FIG. 2. (a) Schematic of our dual-gate BN-encapsulated bilayer $WSe_2$ devices. (b) Photoluminescence (PL) map of a bilayer $WSe_2$ device at varying voltage difference $\Delta V = V_{tg} - V_{bg}$ between the top and bottom gates while keeping the sample charge neutral. The measurements were conducted at sample temperature $T \sim 15$ K under 730-nm continuous laser excitation (incident power ~ 50 μW). (c) Second energy derivative of panel (b). (d) Extracted energy of emission features in (c). The primary QΓ and QK exciton emission ($X_{Q\Gamma}$, $X_{QK}$) and phonon replicas ($X_{Q\Gamma}^{1-5}$, $X_{QK}^{1-6}$) are denoted. The dashed lines are theoretical electric-field-dependent energy shifts of QΓ and QK interband transitions.

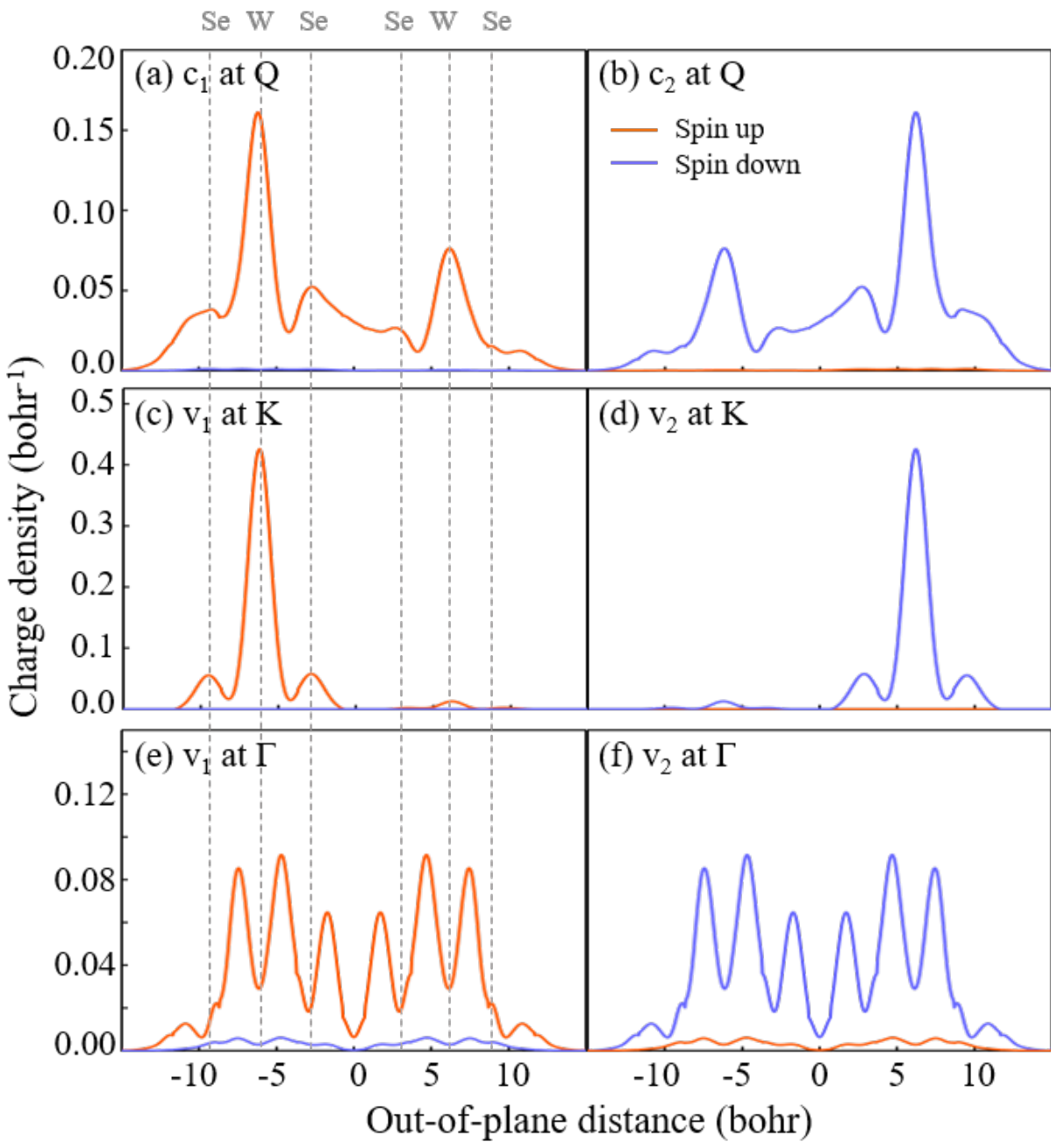

FIG. 3. (a-b) Calculated spin-dependent in-plane-averaged charge density along the out-of-plane direction for the electron states at the Q point of the conduction bands $c_1$ and $c_2$ at an infinitesimal vertical electric field. (c-f) Similar plots for the states at the K and Γ points of the valence bands $v_1$ and $v_2$. The states at $c_1$ ($v_1$) and $c_2$ ($v_2$) have opposite spin and layer polarization. The vertical dashed lines denote the central position of the atoms.

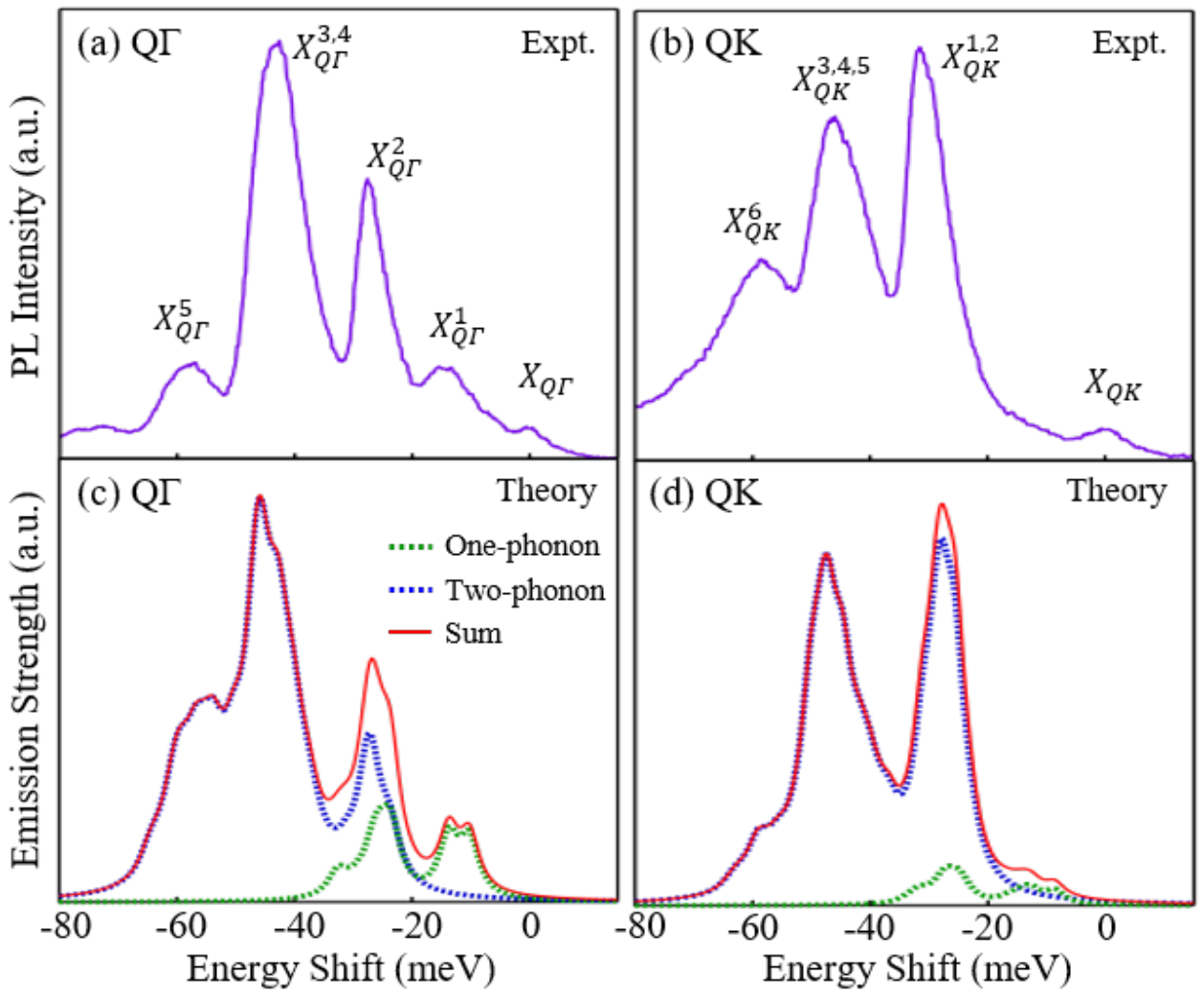

FIG. 4. (a) PL spectrum at $\Delta V = 0$ in Fig. 2(b), offset by the $X_{Q\Gamma}$ energy. (b) PL spectrum at $\Delta V =$ -10 V in Fig. 2(b), offset by the $X_{QK}$ energy. (c-d) Simulated spectra of one-phonon and two-phonon replicas for $X_{Q\Gamma}$ (c) and $X_{QK}$ (d). Both theoretical spectra are broadened for 2 meV to match the experimental peak widths.

**Supplementary Materials of**

# "Electric-field-tunable intervalley excitons and phonon replicas in bilayer $WSe_2$"

Mashael M. Altaiary[1], Erfu Liu[1], Ching-Tarng Liang[2], Fu-Chen Hsiao[2,3], Jeremiah van Baren[1], Takashi Taniguchi[4], Kenji Watanabe[5], Nathaniel M. Gabor[1,6], Yia-Chung Chang[2*], Chun Hung Lui[1*]

[1] Department of Physics and Astronomy, University of California, Riverside, CA 92521, USA.

[2] Research Center for Applied Sciences, Academia Sinica, Taipei 11529, Taiwan.

[3] Advanced Semiconductor Device and Integration Laboratory, Department of Electrical and Computer Engineering, University of Illinois at Urbana-Champaign, Urbana, Illinois 61801, USA

[4] International Center for Materials Nanoarchitectonics (WPI-MANA), National Institute for Materials Science, 1-1 Namiki Tsukuba, Ibaraki 305-0044, Japan.

[5] Research Center for Functional Materials, National Institute for Materials Science, 1-1 Namiki, Tsukuba 305-0044, Japan.

[6] Canadian Institute for Advanced Research, MaRS Centre West Tower, 661 University Avenue, Toronto, Ontario ON M5G 1M1, Canada.

*Corresponding authors. Email: joshua.lui@ucr.edu; yiachang@gate.sinica.edu.tw

## Table of Contents

## 1. Device fabrication and characterization

We fabricate dual-gated bilayer $WSe_2$ devices with hexagonal boron nitride (BN) encapsulation by micro-mechanical exfoliation and co-lamination of thin crystals. We use bulk $WSe_2$ from HQ Graphene Inc.. We first exfoliate bilayer $WSe_2$, thin graphite flakes, and thin BN flakes from their bulk crystals onto silicon substrates with 285-nm-thick $SiO_2$ epilayer. Afterward, we apply a dry-transfer technique to stack the different thin crystals together. We use a polycarbonate stamp to sequentially pick up the top-gate electrode (thin graphite), top-gate dielectric (thin BN), bilayer $WSe_2$, contact electrode (thin graphite), bottom-gate dielectric (thin BN), and bottom-gate electrode (thin graphite). Afterward, we transfer the stack of materials onto a $Si/SiO_2$ substrate and apply the standard electron beam lithography to deposit 100-nm-thick gold electrodes. Finally, the devices are annealed at 360 °C for six hours in an argon environment to cleanse the interfaces.

Fig S1(a) displays the optical image of Device #1, whose results are presented in the main paper. The thickness of the BN flake for the top (bottom) gate of Device #1 is determined to be 26 nm (23.7 nm) by atomic force microscope (AFM).

Fig. S1(b) displays the bottom-gate-dependent photoluminescence (PL) map of bilayer $WSe_2$ with the top gate grounded. We observe the emission of trions (or exciton polarons) on the electron and hole sides. In the main paper, we focus on the exciton PL spectra at the charge neutrality at varying vertical electric field.

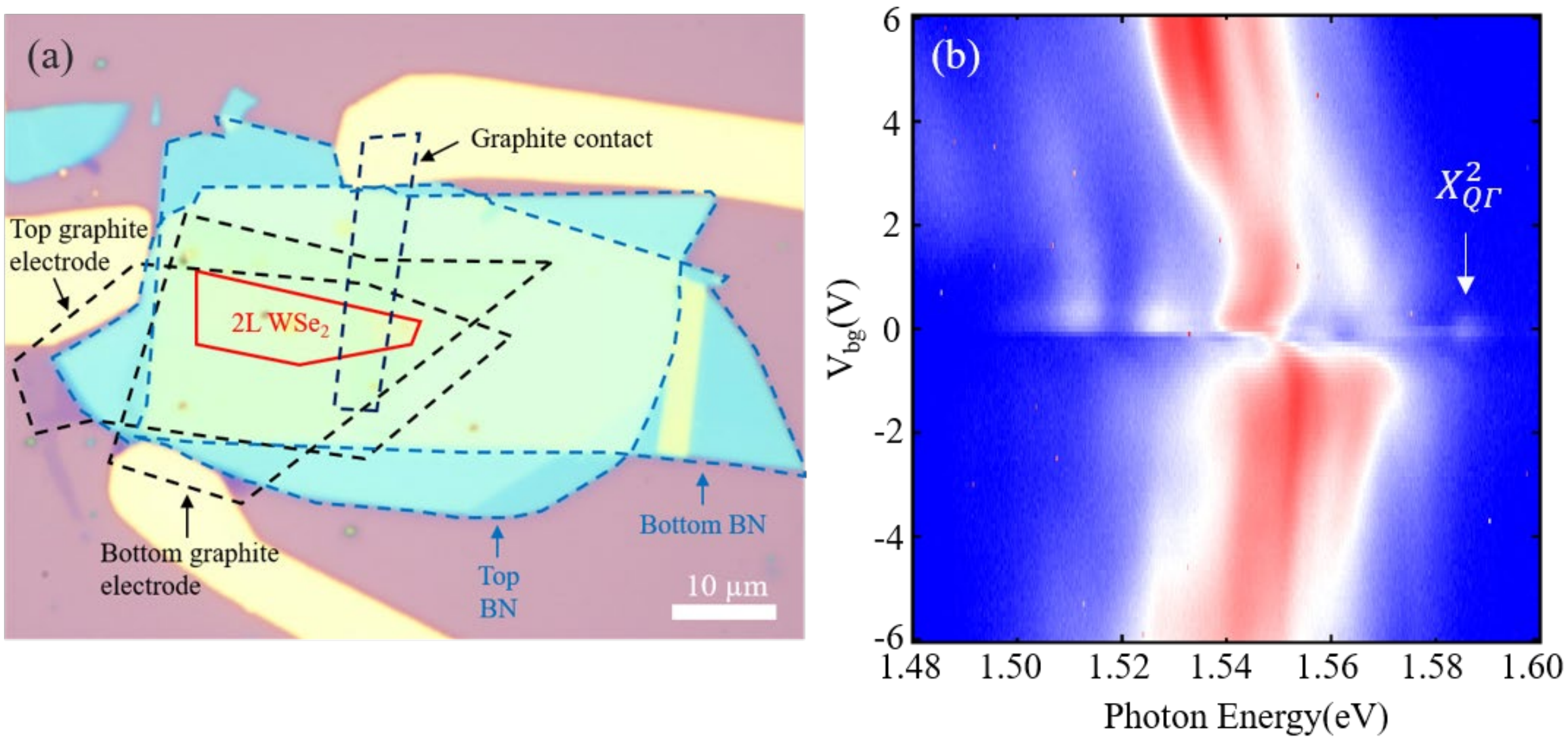

**Fig. S1.** (a) Optical image of BN-encapsulated bilayer $WSe_2$ Device #1 used in the main paper (b) The bottom-gate-dependent photoluminescence (PL) map of the device at sample temperature $T \sim 15$ K under 633-nm continuous laser excitation with incident power of 100 μW. The top gate is grounded in the measurement.

## 2. Determination of the electric field across bilayer $WSe_2$

In our experiment, we have applied gate voltages of opposite signs on the bottom gate ($V_{bg}$) and top gate ($V_{tg}$) to induce a vertical electric field across bilayer $WSe_2$. We adjust the ratio between $V_{tg}$ and $V_{bg}$ so that the sample remains charge neutral in the experiment. We solve the electrostatic problem in our devices to determine the strength of the electric field ($E$) across bilayer $WSe_2$ from $\Delta V = V_{tg} - V_{bg}$.

Fig. S2 displays the electrostatic geometry of our BN-encapsulated devices, which consists of a $WSe_2$ bilayer with thickness $d$ sandwiched between two BN layers with thickness $d_1$ and $d_2$. The static dielectric constant of bilayer $WSe_2$ and BN along the vertical direction are denoted as $\varepsilon_{2L}$ and $\varepsilon_{BN}$, respectively. The electric fields across the top BN, 2L $WSe_2$, and bottom BN are denoted as $E_1$, $E$, $E_2$, respectively. The voltage difference ($\Delta V$) between the top and bottom gates is related to the electric fields as

$$\Delta V = d_1 E_1 + dE + d_2 E_2 \quad (1)$$

As the electrostatic gating injects no net charge into the $WSe_2$ bilayer, the boundary conditions at the two interfaces are

$$\varepsilon_{BN} E_1 = \varepsilon_{2L} E = \varepsilon_{BN} E_2 \quad (2)$$

Combining Eq. (1) and (2), we obtain

$$E = \frac{\Delta V}{\frac{\varepsilon_{2L}}{\varepsilon_{BN}}(d_1 + d_2) + d} \quad (3)$$

In our calculation, we use $d = 1.3$ nm for the $WSe_2$ bilayer, $d_1 = 26$ nm and $d_2 = 23.7$ nm for the top and bottom BN, respectively, for Device #1. We use dielectric constant $\varepsilon_{2L} = 5.51$ for bilayer $WSe_2$, which is obtained from our theoretical calculation in Section 5.1 below. We use dielectric constant $\varepsilon_{BN} = 2.69$ for BN, which is consistent with the literature [1-3]. The calculated electric field for varying $\Delta V$ is shown in Fig. 2(d) of the main paper.

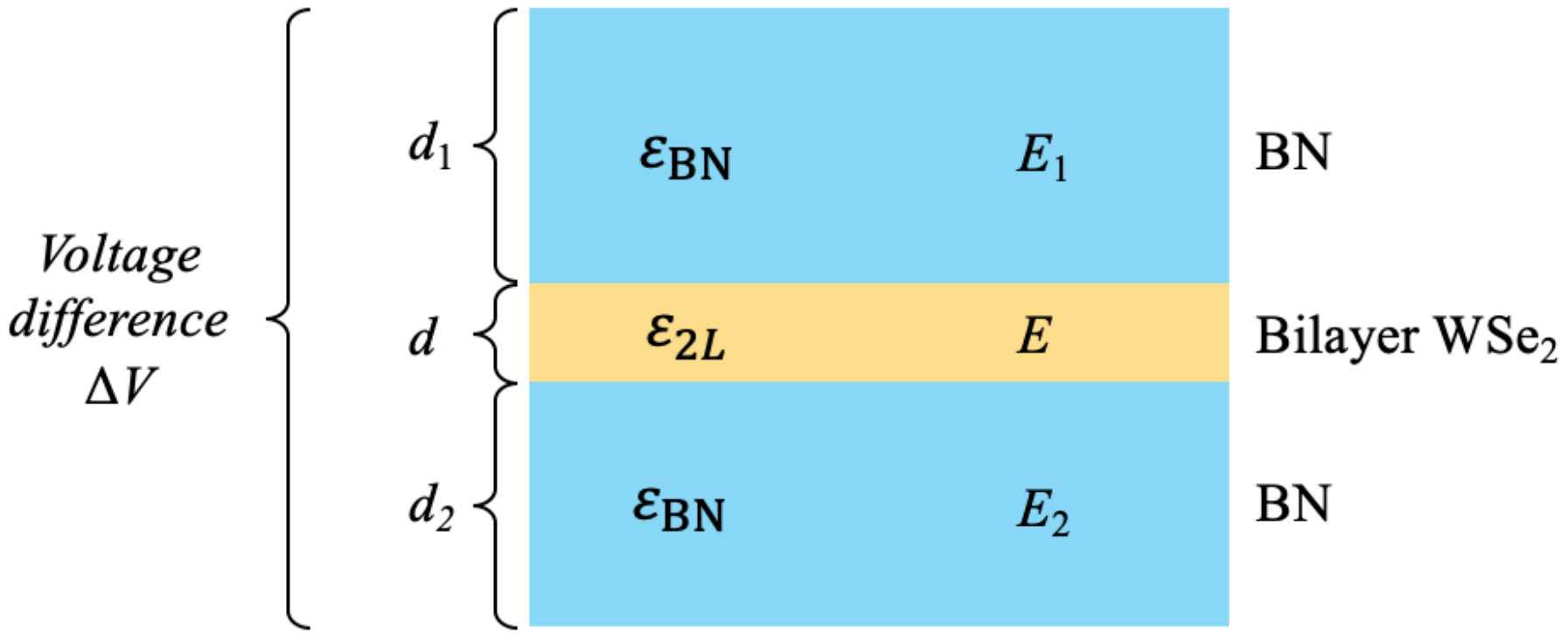

**Fig. S2.** Electrostatic geometry of our BN-encapsulated bilayer $WSe_2$ devices

### 3. Zeeman splitting of the QΓ exciton emission in bilayer $WSe_2$

The QΓ exciton state is formed by the hole states in the Γ valley and the electron states in six energy-degenerate Q valleys. The six Q valleys can be separated into two groups – the $Q_1$, $Q_2$, $Q_3$ valleys and the $Q_1'$, $Q_2'$, $Q_3'$ valleys [see the inset of Fig. 1(b) in the main paper]. The three valleys in each group are related to each other by three-fold rotational symmetry. The two groups of valleys are related to each other by time-reversal symmetry.

The application of a vertical magnetic field preserves the three-fold rotational symmetry but breaks the time-reversal symmetry. As a result, the degeneracy between the $Q_1$, $Q_2$, $Q_3$ valleys and the $Q_1'$, $Q_2'$, $Q_3'$ valleys is lifted. Correspondingly, the QΓ exciton state will be split into two states, each associated with either the $Q_1$, $Q_2$, $Q_3$ valleys or the $Q_1'$, $Q_2'$, $Q_3'$ valleys. Moreover, the two-fold generate states at the Γ point will also be split under magnetic field and contribute to the splitting of the QΓ exciton. Similar valley Zeeman splitting effect has been studied extensively for the K-valley excitons in monolayer TMDs [4-7].

We have measured the PL spectra of QΓ exciton and replicas under finite vertical magnetic field and electric field (Fig. S3). We cannot see clearly the primary $X_{Q\Gamma}$ line due to its weak signal, but we can observe the $X_{Q\Gamma}^2$ replica, which are split into two lines by the Stark effect under electric field. The $X_{Q\Gamma}^2$ replica is further split into four lines under magnetic field. Two lines show right-handed circular polarization and redshift with increasing magnetic field; the other two lines show left-handed circular polarization and blueshift with increasing magnetic field. We attribute such magnetic-field-dependent splitting to the valley Zeeman effect, in which emission with opposite helicity comes from excitons associated with the $Q_1$, $Q_2$, $Q_3$ valleys and the $Q_1'$, $Q_2'$, $Q_3'$ valleys.

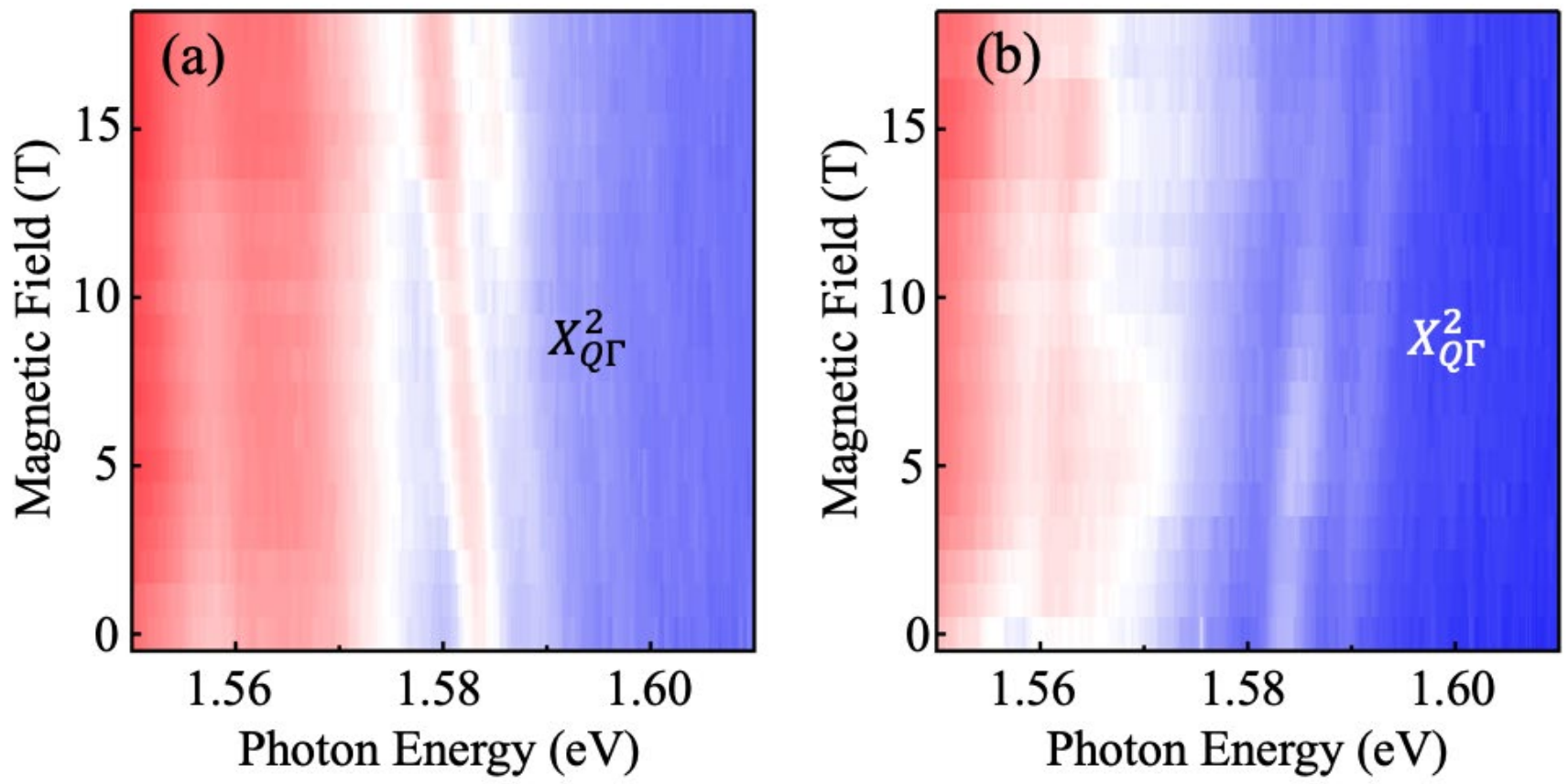

**Fig. S3.** (a-b) Magnetic-field-dependent PL maps of bilayer $WSe_2$ at right-handed (a) and left-handed (b) circular polarization at $\Delta V$ = -1 V under linearly polarized 532-nm laser excitation. The $X_{Q\Gamma}^2$ replica is split into two lines by the Stark effect at finite $\Delta V$. Each line is further split into two lines with opposite helicity and exhibits linear Zeeman shift under magnetic field.

The energy separation ($\Delta E$) between $X_{Q\Gamma}^2$ lines with opposite helicity exhibit linear dependence on the magnetic field, which can be described by $\Delta E = g\mu_B B$, where $\mu_B$ is the Bohr magneton, $B$ is the magnetic field, and $g$ is the effective g-factor. From our data, we extract a g-factor of ~9. This value is consistent with the g-factor (g = 9.5) obtained by a prior experiment that measured quantum-dot emission associated with the QΓ exciton in bilayer $WSe_2$ [8]. Further research is merited to understand the optical selection rules and g-factor of $X_{Q\Gamma}^2$.

## 4. Photoluminescence results of bilayer $WSe_2$ Device #2.

We have obtained similar PL maps in another device (Device #2). The results, presented in Fig. S4, are similar to the results of Device #1 in Fig. 2 of the main paper.

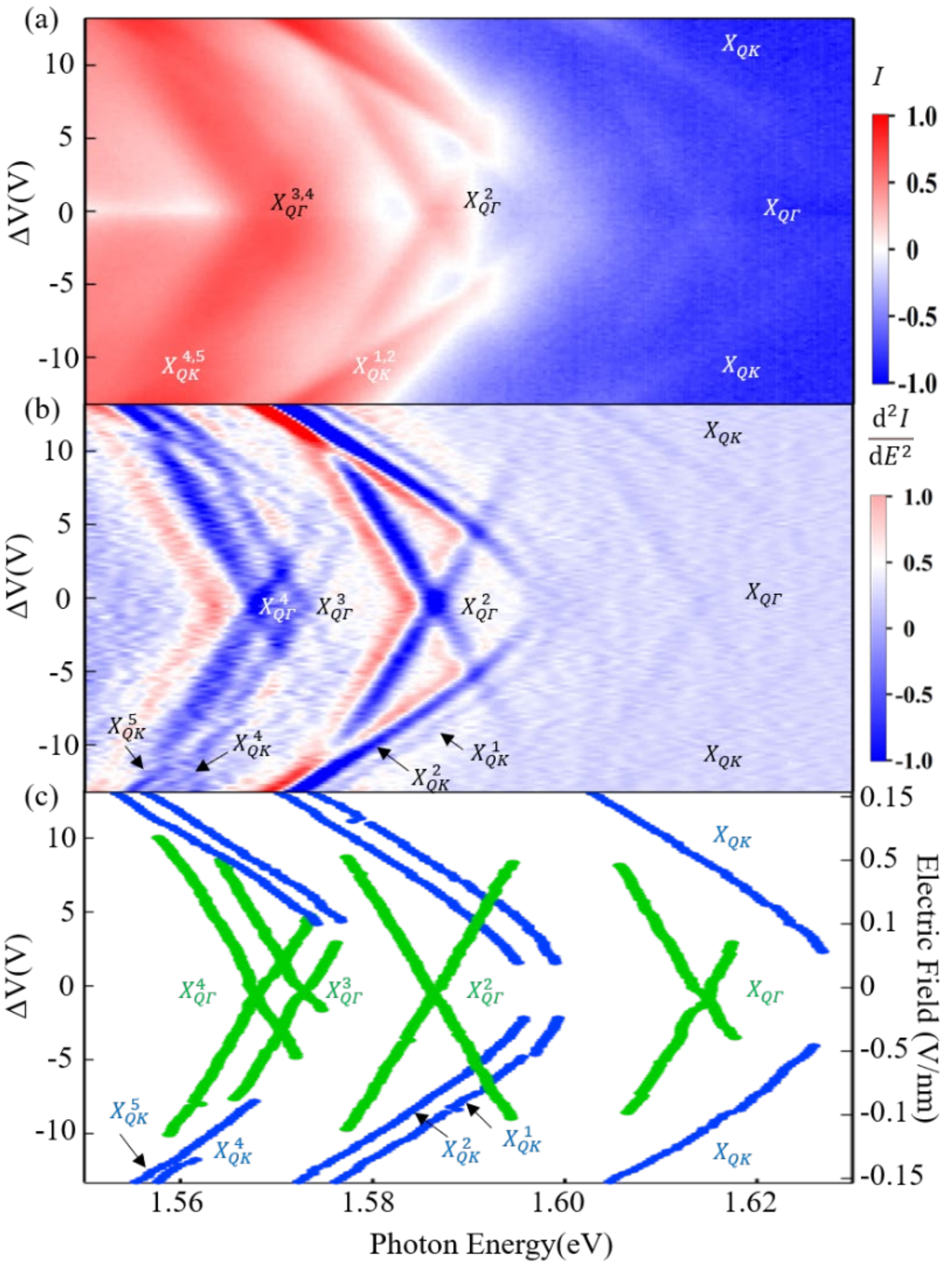

**Fig. S4**. (a) Gate-dependent photoluminescence (PL) map of bilayer $WSe_2$ Device #2. (c) The second energy derivative of panel (a). (c) The extracted peak energy of emission features in (a, b). The measurements were conducted at sample temperature $T \sim 15$ K under 532-nm continuous laser excitation with incident power ~50 μW.

## 5. Theoretical calculations

### 5.1 Electric-field effect on the band energies of bilayer $WSe_2$

We have calculated the shift of band-edge energies and transition energies of various intervalley excitons of bilayer $WSe_2$ under the application of an electric field ($E$) along the $z$-axis (perpendicular to the plane). The applied electric field gives rise to a perturbation term $\widehat{H}_1 = eEz$, which is included in the density-functional theory (DFT) to calculate the effect of electric field on the band structure. The calculation is carried out by using the WIEN2K package [9]. We adopted the generalized gradient approximation (GGA) with a sufficiently large supercell of size $L_z = 5.21$ nm along the $z$-axis. We have included the spin-orbit interaction and the van der Waals interaction between the two $WSe_2$ layers in our calculation.

We first calculate the band structure of unstrained bilayer $WSe_2$ by adopting a lattice constant of 0.328 nm from the experimental value [10]. The calculated band structure displays the conduction band minimum (CBM) at the Q point (near the middle point between K and Γ) and the valance band maximum (VBM) at the K point (Fig. S1). The energy of the top valence band (v1) at the Γ point is 132 meV below that of the VBM at the K point. Such an energy separation is smaller than the Γ-K separation of 140 meV measured by a recent experiment of angle-resolved photoemission spectroscopy (ARPES) on a BN-encapsulated bilayer $WSe_2$ sample [11]. This slight deviation suggests that the BN-encapsulated $WSe_2$ may be slightly strained. Indeed, we find that, by reducing the lattice constant of bilayer $WSe_2$ for 0.098%, our calculated band structure [Fig. S1(b)] can exhibit a Γ-K separation of 140 meV, which matches the ARPES result. Therefore, in all of the following calculations, we consider strained bilayer $WSe_2$ with 0.098% reduced lattice constant.

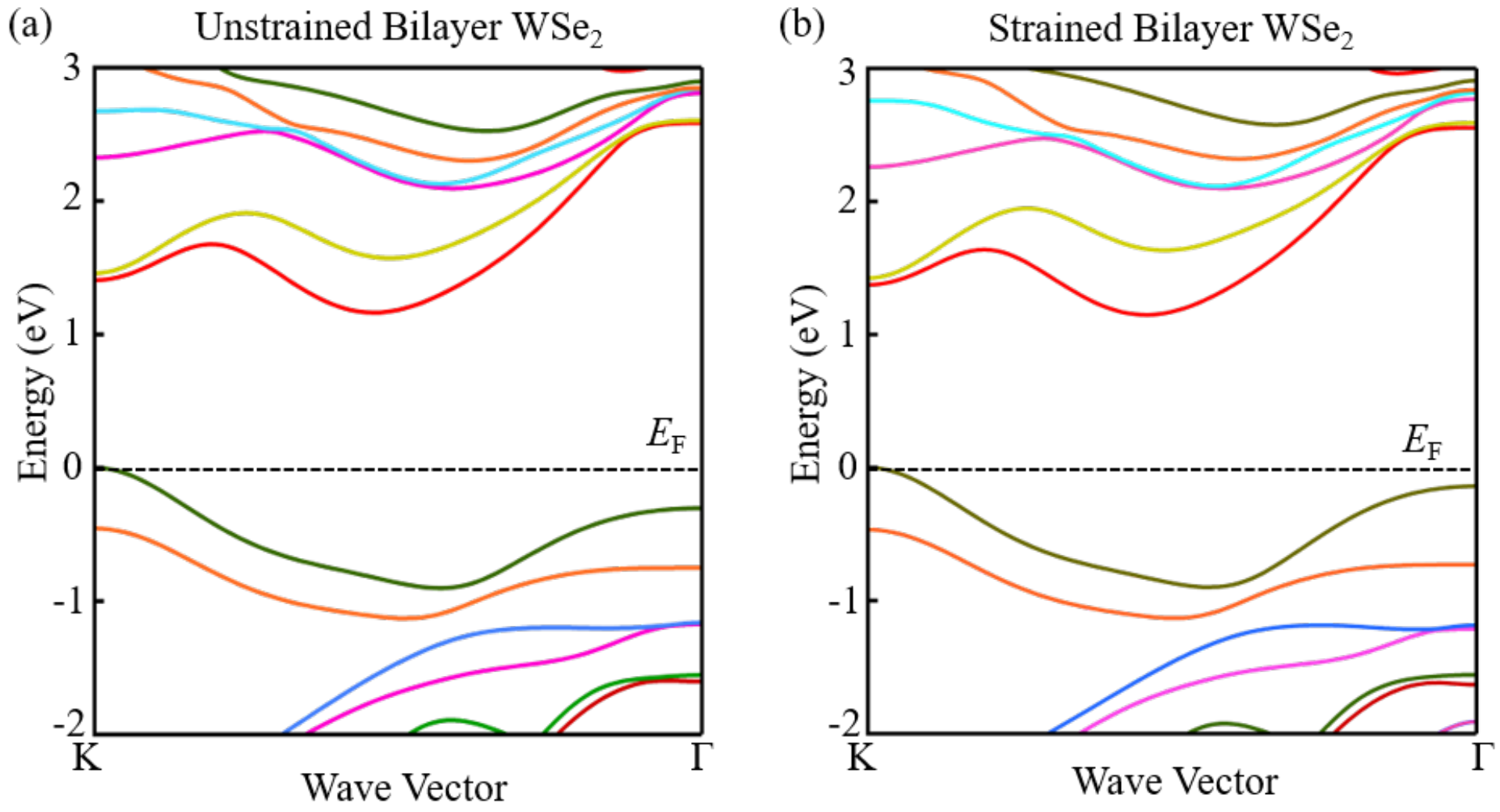

**Fig. S5**. (a) Calculated band structure of unstrained bilayer $WSe_2$. (b) Calculated band structure of strained bilayer $WSe_2$ with 0.098% reduced lattice constant. The calculations are done by the WIEN2K package with GGA.

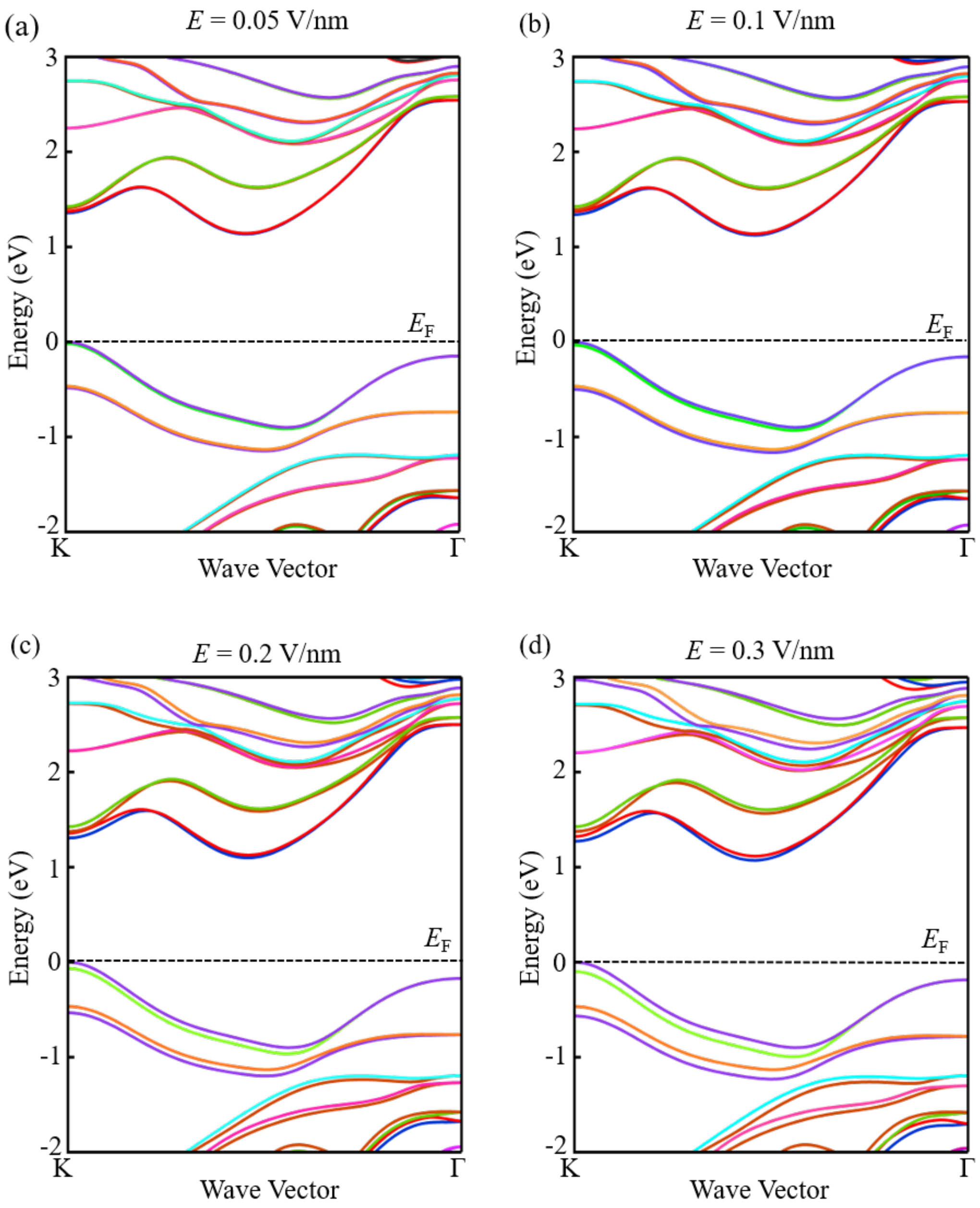

**Fig. S6**. Band structure of bilayer $WSe_2$ with 0.098% reduced lattice constant calculated by using WIEN2k under four different vertical electric fields: (a) $E$ = 0.05 V/nm, (b) $E$ = 0.1 V/nm, (c) $E$ = 0.2 V/nm, and (d) $E$ = 0.3 V/nm.

We then consider the effect of a perpendicular electric field on the strained bilayer $WSe_2$. Fig. S6 displays the calculated band structure of the strained bilayer $WSe_2$ under perpendicular electric fields $E$ = 0.05, 0.1, 0.2, and 0.3 V/nm. These electric fields are the net fields seen by an electron in the material under the screening of the valence electrons. Our results show that the two-fold degenerate conduction band at Q and valence band at K split because the applied electric field breaks the inversion symmetry of bilayer $WSe_2$. However, the valence band at Γ remains two-fold degenerate due to the time-reversal symmetry (Kramer's degeneracy).

In our consideration below, we will focus on the conduction band at the Q point and the valence band at the K and Γ points. We denote the conduction band with lower (higher) energy at Q as c1 (c2), and the valence band with higher (lower) energy at K as v1 (v2). Fig. S7 displays the calculated interband transition energies of the $Q_{c1} - K_{v1}$, $Q_{c1} - \Gamma_{v1}$, and $Q_{c2} - \Gamma_{v2}$ transitions as a function of screened electric field. The $Q_{c1} - \Gamma_{v1}$ and $Q_{c2} - \Gamma_{v2}$ transition energies are found to split with increasing electric field. Here we have not considered the excitonic effect, which should only produce a constant energy offset with insignificant influence on the field-dependent energy shift. In Fig. S7, the Q-Γ transition energies are the interband transition energies obtained by DFT, whereas the Q-K transition energy is shifted to match the experimental separation between the QΓ and QK exciton emission lines at zero field in Fig. 2(d).

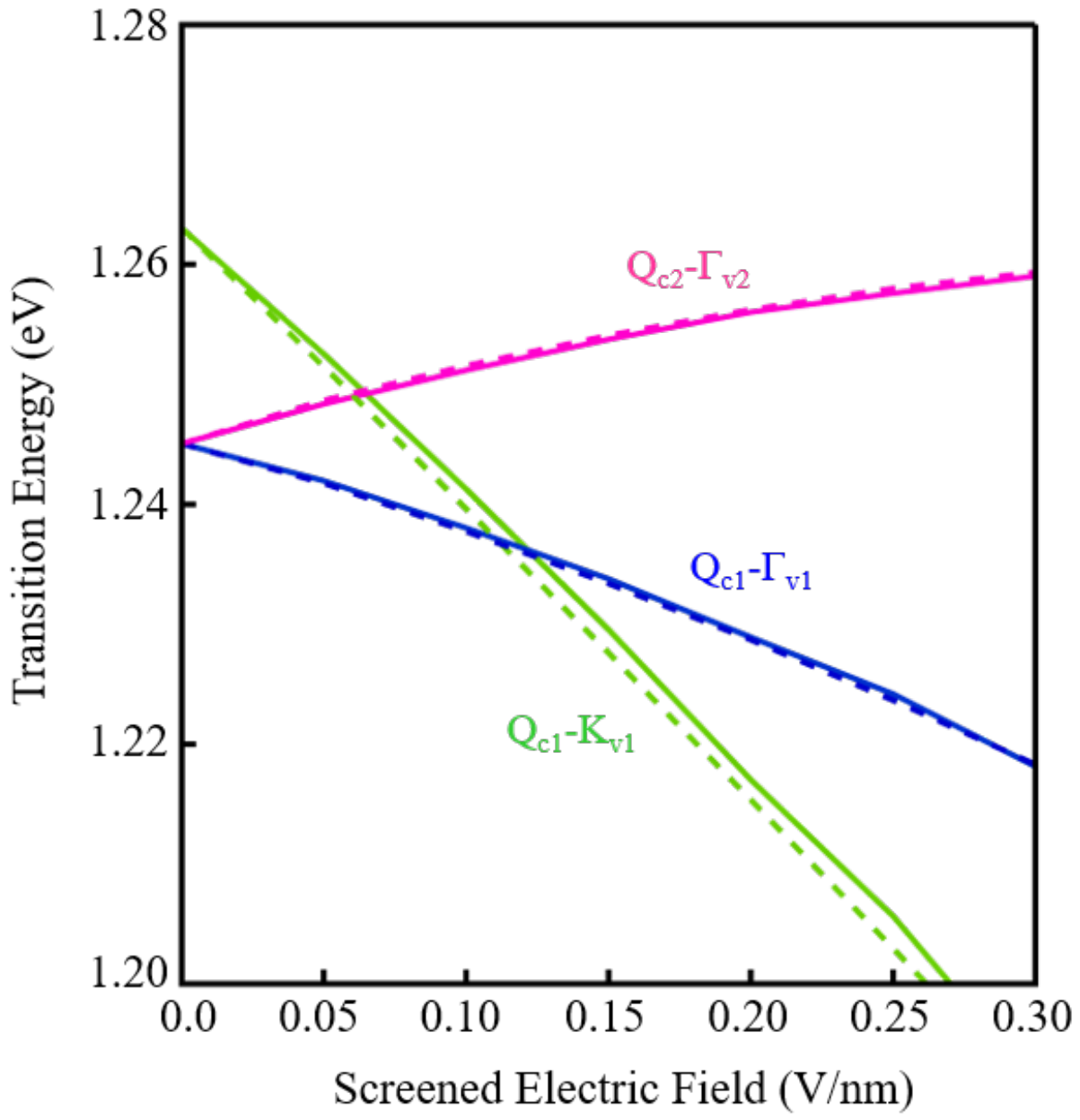

**Fig. S7.** Calculated interband transition energy for the $Q_{c1} - K_{v1}$, $Q_{c1} - \Gamma_{v1}$, and $Q_{c2} - \Gamma_{v2}$ transitions as a function of screened electric field. The solid (dashed) curves are results of self-consistent (non-self-consistent) calculation as described in the text. Here the Q-Γ transition energies are the interband transition energies obtained by DFT, whereas the Q-K transition energy is shifted to match the experimental separation between the QΓ and QK excitons in Fig. 2(d).

We have used two different approaches to calculate the band energy shift due to electric field. In the first approach, we carry out self-consistent calculation, which considers the response of valence electrons and the atomic relaxation to an external electric field (*i.e.* the screening effect); we obtain the band energy shift as a function of external electric field. In the second approach, we carry out non-self-consistent calculation, in which the applied electric field is taken as the net field after all the screening effects. Afterward, we compare the result of the two calculations and find that their electric-field dependence is off by a ratio of 5.51; this ratio is interpreted as the dielectric constant of bilayer $WSe_2$ (i.e. $\varepsilon_{2L}$ = 5.51). Fig. S7 shows the calculated transition energy as a function of screened electric field for both calculations (solid and dashed lines). They appear to match each other well after we use the dielectric constant $\varepsilon_{2L}$ = 5.51 to convert the applied field into screened field in the self-consistent result.

We have used the self-consistent results to compare with our experimental data in Fig. 2(d) of the main paper. We find decent agreement on the field-dependent energy shift between experiment and theory. We note that the absolute transition energies in our calculation differ from the experimental emission energy because DFT cannot predict the band gap accurately and we have not considered the excitonic effect. But such deviation should have no significant influence on the field-dependent energy shift. So, in our comparison, we shift the energies of the QΓ and QK excitons to match the experiment energies at zero electric field and focus our comparison on the field-dependent shift of the transition energy.

## 5.2 Comparison of QΓ and QK exciton emission energies

From the theory-experiment comparison of the electric-field-dependent energy shift [Fig. 2(d)], we can identify the higher-energy emission lines at 1.62 – 1.63 eV as the QK exciton and the lower-energy emission lines at 1.605 – 1.62 eV as the QΓ exciton. However, in the band structure obtained by DFT, the valence band is 140 meV higher at the K point than at the Γ point. That is, the Q-Γ transition energy exceeds the Q-K transition energy by 140 meV. This apparently contradicts against our result that QΓ exciton lies below the QK exciton. This implies that QΓ excitons has much larger binding energy than the QK exciton. There are three factors that contribute to the stronger binding of the QΓ exciton.

First, the QΓ exciton has larger effective mass than the QK exciton. Table S1 lists the effective carrier mass at the K, Γ, and Q points calculated by WIEN2k. From the orientation-averaged effective mass at the K, Γ, and Q points, we deduce the reduced mass ($\mu_X$) for the QK and QΓ excitons to be $0.247m_0$ and $0.417m_0$ respectively, where $m_0$ is the free electron mass. If we assume that the QK exciton binding energy is ~150 meV, then the QΓ exciton binding energy can reach ~253 meV by considering their ratio of effective mass. Such enhanced binding energy can partially counter the 140 meV difference in the QΓ ad QK interband transition energies.

Second, in the Brillouin zone of bilayer $WSe_2$, there are six inequivalent Q valleys, all of which contribute equally to the formation of QΓ exciton. In contrast, only three of the six Q valleys contribute significantly to the QK exciton formation, whereas the other three are further away and have negligible contribution. The double number of Q valleys in the QΓ exciton formation can significantly enhance its binding energy compared to the QK exciton.

Third, the larger effective mass of QΓ exciton can cause a larger spread of the exciton envelop function in the k-space, which can lead to significant intervalley mixing and further enhancement of the excitonic binding. For instance, previous studies of donor states in silicon shows that the intervalley mixing effect can increase the donor binding energy by 50% [12]. Similar effect is expected in the QΓ intervalley exciton here.

By combining the three factors above, it is reasonable that the QΓ exciton binding energy can exceed the QK exciton binding energy by more than 140 meV, thus making the QΓ exciton emission energy lower than the QK exciton energy.

| | $K_{v1}$ | $\Gamma_{v1}$ | $Q_{v1}$ | $Q_{c1}$ | $K_{c1}$ | $K_{c2}$ |
|---|---|---|---|---|---|---|
| $m_{\|\|}$ | 0.42 | 1.47 | ∞ | 0.49 | 0.44 | 0.34 |
| $m_{\perp}$ | 0.44 | 1.49 | 1.47 | 0.72 | 0.43 | 0.32 |

**Table S1.** Band-edge effective masses (in unit of $m_0$) at K, Γ, and Q points. We note that the $Q_{v1}$ band has a linear dispersion along the Γ-K axis at the Q point; thus, we take $1/m_{||} = 0$.

## 5.3 Density distribution and electric dipole of band-edge states in bilayer $WSe_2$

To understand the physical mechanism for the electric-field-induced energy splitting, we calculate the *z*-dependent charge density averaged over the *x*-*y* plane for the Bloch states in the bottom conduction band (c1) at Q and the top valence band (v1) at K and Γ in bilayer $WSe_2$ via DFT by using the relation

$$\rho_\mu(z_e) = \left\langle \psi_{\mu,K} \middle| \frac{1}{L_c} \sum_{g_z} e^{ig_z(z-z_e)} \middle| \psi_{\mu,K} \right\rangle, \quad (4)$$

where μ labels the Block states ($K_{v1}$, $\Gamma_{v1}$, $Q_{c1}$, and $Q_{c2}$) of concern. Since $\rho_\mu$ is not provided by the WIEN2k package, we use the wave functions obtained from the LASTO (linear augmented Slater-type orbitals) package [13-15] to calculate the in-plane averaged charge density according to Eq. (4). Here the Bloch states are calculated via a supercell method and $L_c$ is the length of the supercell used. The band structures obtained by LASTO package are nearly the same as those by WIEN2k with suitable choice of LASTO basis functions. Both approaches are based on all-electron (both core and valence electrons) DFT with augmented basis functions. The LASTO basis functions are properly selected linear combinations of the augmented plane waves (APWs).

Fig. 3 in the main paper displays our calculated charge densities $\rho_\mu(z)$ for the $K_{v1}$, $\Gamma_{v1}$, $Q_{c1}$, and $Q_{c2}$ states in bilayer $WSe_2$ under a small electric field ($E$ = 0.001 V/nm). The purpose of using a small electric field is to break the inversion symmetry so that we can see the spin and layer dependence of the wavefunctions when their degeneracy is lifted. The results allow us to calculate the dipole moment and the spin angular momentum of each state. The calculated dipole moment for the $K_{v1}$, $\Gamma_{v1}$, and $Q_{c1}$ states are 5.81, 0, and 2.02 atomic unit (a.u.), respectively.

## 5.4 Calculation of phonon modes by a rigid-ion model for bilayer $WSe_2$

Besides the field-dependent emission energy shift discussed above, our experiment also reveals phonon replicas of the QΓ and QK excitons. We have carried out comprehensive theoretical calculations on the phonon replica spectra. In this section, we will first describe our calculation of the phonon band structure and phonon polarization vectors of bilayer $WSe_2$. In the following sections, we will present our theory to calculate the phonon replica spectra.

We adopt a rigid-ion model (RIM) [16] to obtain the phonon polarization vectors. In this model, we consider only the short-range ion-ion Coulomb interaction. We neglect the long-range ion-ion Coulomb interaction (due to charge transfer between W and Se atoms), which should be insignificant given the very small energy splitting between the LO and TO phonon modes at the zone center [17]. The weak long-range ion-ion interaction indicates that $WSe_2$ is a highly covalent material with little inter-atomic charge transfer.

Fig. S8 displays the configuration and labels of the atoms in the bilayer $WSe_2$ unit cell, which includes one W atom and two Se atoms in both the bottom and top $WSe_2$ monolayers. In the unit cell, we define the position of the W atom in the bottom $WSe_2$ monolayer as the origin of our coordinate system. The two nearby Se atoms are located at $(a/\sqrt{3}, 0, \pm d/2)$, where $a$ is the in-plane lattice constant and $d$ is the vertical distance between the lower and upper Se atoms in the bottom $WSe_2$ monolayer. Below we will define a set of dynamic matrices to obtain the phonon dispersion. The form of these matrices is determined by symmetry analysis.

The dynamic matrix for the nearest-neighbor interaction between the W atom at the origin (labeled as 1) and the Se atom at $(a/\sqrt{3}, 0, -d/2)$ (labeled as 2) has the form of

$$D^{12} = \begin{pmatrix} A_1 & 0 & D_1 \\ 0 & B_1 & 0 \\ E_1 & 0 & C_1 \end{pmatrix},$$

where $A_1 - E_1$ denote the interaction parameters for the nearest-neighbor interaction.

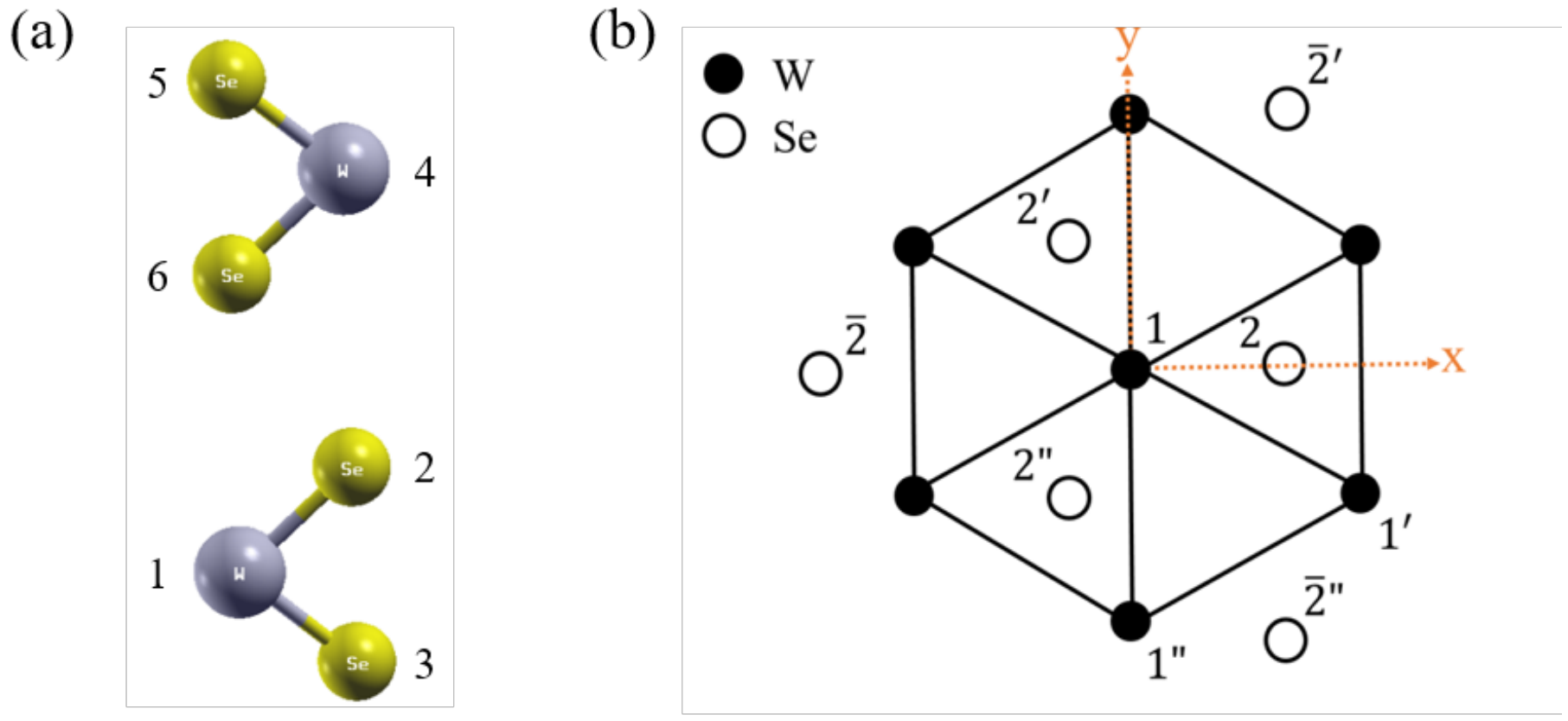

**Fig. S8.** (a) The side view of the unit cell of bilayer $WSe_2$. (b) The bottom view of the lower $WSe_2$ layer in the unit cell. The numerical labels for different atoms are shown.

The dynamic matrix between the Se atom at $(a/\sqrt{3}, 0, -d/2)$ (labeled as 2) and the Se atom at $(-a/\sqrt{3}, -a, -d/2)$ (labeled as $2'$) has the form:

$$D^{22'} = \begin{pmatrix} A_2 & 0 & D_2 \\ 0 & B_2 & E_2 \\ D_2 & -E_2 & C_2 \end{pmatrix}.$$

The dynamic matrix between the W atom at the origin (labeled by 1) and the W atom at $(0, -a, 0)$ (labeled by $1'$) has the form:

$$D^{11'} = \begin{pmatrix} A_3 & 0 & 0 \\ 0 & B_3 & 0 \\ 0 & 0 & C_3 \end{pmatrix}.$$

The dynamic matrix for the next-neighbor interaction between the Se atom at $(a/\sqrt{3}, 0, -d/2)$ (labeled as 2) and the Se atom at $(a/\sqrt{3}, 0, d/2)$ has the form:

$$D^{23} = \begin{pmatrix} A_4 & 0 & 0 \\ 0 & A_4 & 0 \\ 0 & 0 & C_4 \end{pmatrix}.$$

The dynamic matrix for the second-nearest-neighbor interaction between the W atom at the origin (labeled as 1) and the Se atom at $(-2a/\sqrt{3}, 0, -d/2)$ (labeled as $\bar{2}$) has the form:

$$\mathrm{D}^{1\bar{2}} = \begin{pmatrix} A_5 & 0 & D_5 \\ 0 & B_5 & 0 \\ E_5 & 0 & C_5 \end{pmatrix}.$$

The above five dynamic matrices describe the phonons of monolayer $WSe_2$. In bilayer $WSe_2$, a second $WSe_2$ layer sits on top of the bottom $WSe_2$ layer according to the 2H stacking order [Fig. S8(a)]. In the top $WSe_2$ monolayer, we label the W atom as 4, the lower (upper) Se atom as 6 (5) [Fig. S8(a)]. The dynamic matric for the interlayer interaction between Se atom 6 and Se atom 2 has the form:

$$D^{26} = \begin{pmatrix} A_6 & 0 & 0 \\ 0 & B_6 & 0 \\ 0 & 0 & C_6 \end{pmatrix}.$$

The dynamic matrices between other atoms can be related to the above six matrices by symmetry. We have adjusted the interaction parameters in these dynamic matrices to fit the DFT results of phonon dispersion [17] and the phonon replica PL data of monolayer $WSe_2$ [18,19]. The best-fit force constants are listed in Table S2. Afterward, we use these force constants to generate the bilayer phonon dispersion and polarization vector in our rigid-ion model. The calculated phonon band structure is presented in Fig. S9(a). The polarization vectors of several relevant phonon modes at the M, K, and Q points are shown in Fig. S9(b).

| | | | | | | | | | | | |
|---|---|---|---|---|---|---|---|---|---|---|---|
| $A_1$ | $B_1$ | $C_1$ | $D_1$ | $E_1$ | $A_2$ | $B_2$ | $C_2$ | $D_2$ | $E_2$ | $A_3$ | $B_3$ |
| -10.306 | -1.276 | -9.823 | 7.901 | 7.105 | -0.207 | -1.849 | 0.835 | 0.158 | -0.152 | -1.712 | 0.078 |
| $C_3$ | $A_4$ | $C_4$ | $A_5$ | $B_5$ | $C_5$ | $D_5$ | $E_5$ | $A_6$ | $B_6$ | $C_6$ | |
| 1.375 | 0.520 | -3.487 | -0.238 | -1.412 | -0.330 | 1.019 | 0.160 | -0.0758 | -0.0758 | -0.19 | |

**Table S2.** Interaction parameters (in unit of $2e^2/a^2d$) used in our rigid-ion model for bilayer $WSe_2$. $a$ and $d$ are defined in the text.

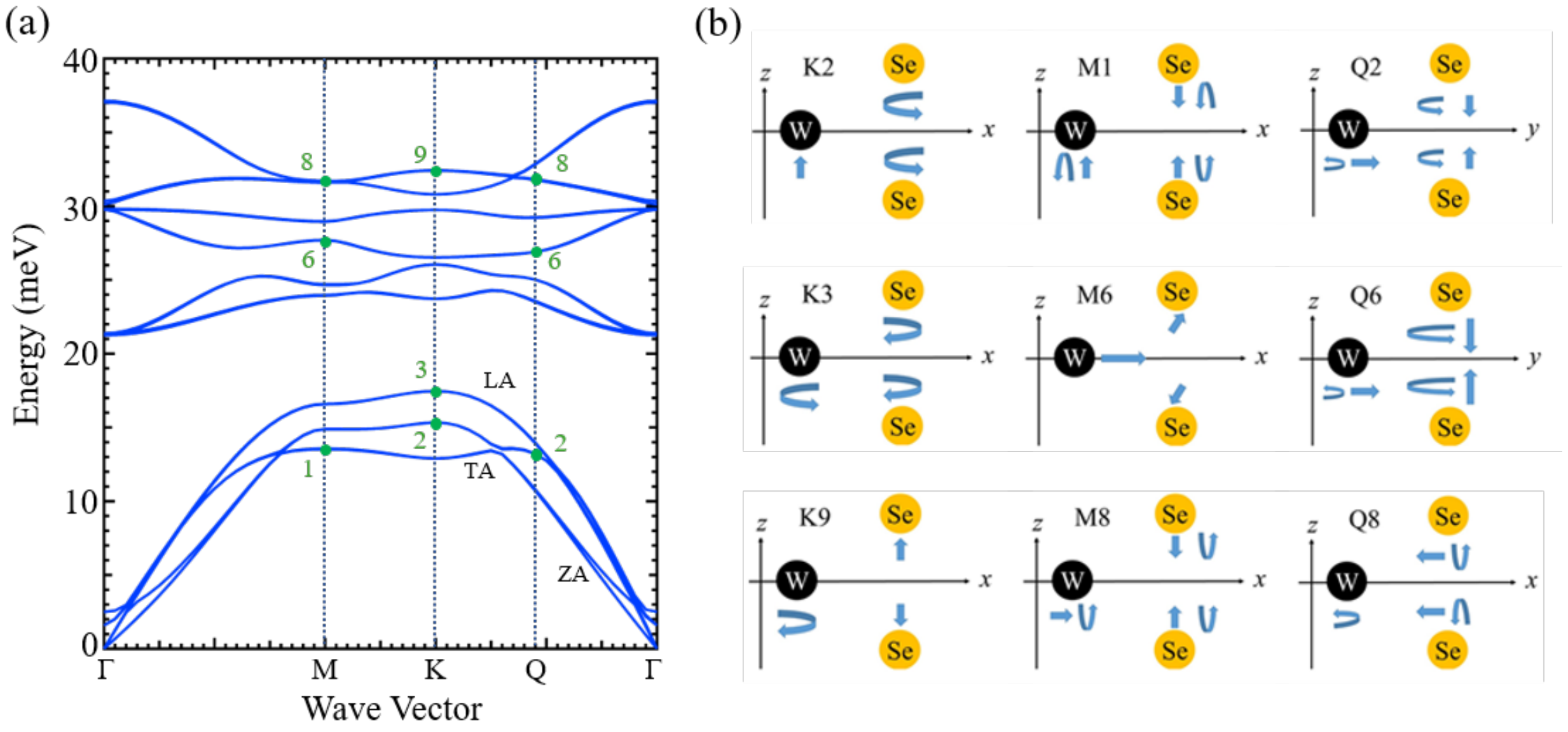

**Fig. S9.** (a) Phonon dispersion of bilayer $WSe_2$ calculated by our rigid-ion model. There are nine pairs of phonon branches (labeled as 1-9 from low to high energy); each pair is nearly degenerate except the interlayer shear and breathing modes. (b) Schematic polarization vectors of representative phonon modes at the K, M, Q points. Corresponding labels of the phonon modes are denoted in both panels (a) and (b). The straight (curved) arrows denote linear (circular) polarization. The arrow size is proportional to the displacement magnitude. Some atoms at the M and Q points have both linear and chiral polarization components. For simplicity, we only show the polarization configurations of the lower $WSe_2$ layer, while the configurations of the upper layer are related by inversion symmetry.

### 5.5 Calculation of one-phonon replica spectra

Besides the field-dependent emission energy shift discussed above, our experiment also reveal phonon replica of the QΓ and QK excitons, which exhibit the same field dependence but are redshifted from the primary emission lines. Here we will present our theoretical calculation of the photon replica spectra.

The phonon replicas can come from one-phonon or multi-phonon emission processes. We first consider the one-phonon process. The emission strength of one-phonon-assisted intervalley exciton recomination is proportional to the transition rate ($P^{(1)}$) according to Fermi's golden rule and the perturbation theory:

$$P_\mu^{(1)}(\omega) = \frac{2\pi}{\hbar}\sum_\alpha \left|\sum_\nu \frac{\langle \omega;\ \Omega_\alpha|\hat{H}_{el}|X_\nu;\ \Omega_\alpha\rangle\langle X_\nu;\Omega_\alpha|\hat{H}_{ep}|X_\mu\rangle}{E_\mu - E_\nu - \hbar\Omega_\alpha + i\gamma}\right|^2 \delta(E_\mu - \hbar\Omega_\alpha - \hbar\omega). \quad (5)$$

Here $\hat{H}_{el}$ ($\hat{H}_{ep}$) is the electron-light (electron-phonon) interaction Hamiltonian; ω denotes the frequency of emitted photon; $\Omega_\alpha$ denotes the frequency of the emitted phonon in mode α; $X_\mu$ with μ = QΓ or QK denotes the initial intervalley exciton state with energy $E_\mu$; $X_\nu$ denotes the mediating exciton state with energy $E_\nu$ after the emission of one phonon; γ = 2 meV is a phenomenological broadening energy due to finite carrier lifetime; The associated exciton-phonon scattering strength is defined as:

$$S_{\nu\mu}^\alpha = N\left|\frac{\langle X_\nu;\Omega_\alpha|\hat{H}_{ep}|X_\mu\rangle}{E_\mu - E_\nu - \hbar\Omega_\alpha + i\gamma}\right|^2. \quad (6)$$

Here $N$ is the number of unit cells in the sample. For bilayer $WSe_2$, each electronic band is doubly degenerate at zero electric field; here we have summed over the contributions of the two degenerate states associated with level $\nu$. Moreover, bilayer $WSe_2$ shows nine pairs of phonon branches, which are here labeled by 1 – 9 from low to high energy; each pair is nearly degenerate (except the interlayer shear and breathing modes) due to the weak interlayer lattice coupling [Fig. S9(a)]. We have summed over the contributions from the two nearly degenerate phonon modes in our results of $S_{\nu\mu}^\alpha$.

With $S_{\nu\mu}^\alpha$ defined in Eq. (6), the transition rate in Eq. (5) can be simplified as:

$$P_\mu^{(1)}(\omega) = \frac{2\pi}{N\hbar}\sum_{\alpha\nu}\left|\langle\omega|\hat{H}_{el}|X_\nu\rangle\right|^2 S_{\nu\mu}^\alpha \delta(E_\mu - \hbar\Omega_\alpha - \hbar\omega). \quad (7)$$

To obtain $P_\mu^{(1)}$, we need to first calculate $S_{\mu\nu}^\alpha$ for the relevant mediating states. We note that the mediating $X_\nu$ states don't need to bound excitonic states, but they must be momentum-direct excitonic states to have finite optical transitions.

For the $X_{QK}$ initial state, the possible mediating $X_\nu$ states include the direct excitonic states in the K valley or Q valley (denoted as $X_{KK}$ and $X_{QQ}$ here). For the $X_{Q\Gamma}$ initial state, the possible mediating states $X_\nu$ include $X_{QQ}$ and $X_{\Gamma\Gamma}$. Here K and Q denote any of the two K points and six Q points; the former (latter) subscript refers to the electron (hole) state. We have calculated $S_{\mu\nu}^\alpha$ by DFT for different exciton-phonon scattering processes that emit a phonon. The results are presented in Table S3. The calculated $S_{\nu\mu}^\alpha$ values are overall quite small, because all of these one-phonon scattering processes are non-resonant. For all of the listed scattering processes for QΓ and QK initial excitons, the emitted phonons are near the Q, M, and K points (and their symmetry-related points).

| Exction-phonon transitions | Phonon momentum | Phonon branch pairs from low to high energy | | | | | | | | |
|---|---|---|---|---|---|---|---|---|---|---|
| | | 1 | 2 | 3 | 4 | 5 | 6 | 7 | 8 | 9 |
| $X_{Q_1'K} \to X_{KK}$<br>$X_{Q_2'K} \to X_{KK}$<br>$X_{Q_3'K} \to X_{KK}$ | M | 0.093<br>(13.7) | 0. 080<br>(14.9) | 0.053<br>(16.8) | 0.021<br>(24.1) | **0.221**<br>**(24.7)** | **0.113**<br>**(27.7)** | 0.033<br>(29.0) | 0.087<br>(31.6) | 0.078<br>(33.7) |
| $X_{Q_1K} \to X_{KK}$<br>$X_{Q_2K} \to X_{KK}$<br>$X_{Q_3K} \to X_{KK}$ | Q | **0.237**<br>**(8.57)** | 0.151<br>(12.3) | 0.082<br>(12.6) | 0.028<br>(23.1) | **0.458**<br>**(25.0)** | **0.511**<br>**(26.8)** | 0.045<br>(29.4) | 0.015<br>(31.9) | 0.052<br>(33.7) |
| $X_{Q_1K} \to X_{Q_1Q_1}$<br>$X_{Q_2K} \to X_{Q_2Q_2}$<br>$X_{Q_3K} \to X_{Q_3Q_3}$ | Q | **0.172**<br>**(8.71)** | 0.030<br>(12.5) | 0.066<br>(12.8) | 0.013<br>(23.1) | 0.034<br>(25.0) | 0.050<br>(26.8) | 0.015<br>(29.4) | 0.084<br>(31.9) | 0.034<br>(33.7) |
| $X_{Q_1'K} \to X_{Q_1'Q_1'}$<br>$X_{Q_2'K} \to X_{Q_2'Q_2'}$<br>$X_{Q_3'K} \to X_{Q_3'Q_3'}$ | M | **0.102**<br>**(13.7)** | 0.062<br>(14.9) | 0.060<br>(16.8) | 0.025<br>(24.1) | 0.036<br>(24.7) | **0.204**<br>**(27.7)** | 0.021<br>(29.0) | 0.048<br>(31.6) | 0.034<br>(31.7) |
| $X_{Q\Gamma} \to X_{QQ}$<br>for any Q valley | Q | **0.157**<br>**(10.1)** | **0.312**<br>**(13.5)** | **0.294**<br>**(14.0)** | 0.077<br>(23.5) | 0.045<br>(25.5) | **0.301**<br>**(26.6)** | 0.048<br>(29.3) | 0.077<br>(32.0) | **0.128**<br>**(32.8)** |
| $X_{Q\Gamma} \to X_{\Gamma\Gamma}$<br>for any Q valley | Q | **0.313**<br>**(10.1)** | 0.012<br>(13.5) | 0.010<br>(14.0) | **0.433**<br>**(23.5)** | **0.201**<br>**(25.5)** | 0.039<br>(26.6) | 0.029<br>(29.3) | 0.021<br>(32.0) | 0.030<br>(32.8) |

**Table S3.** The $S_{\nu\mu}^{\alpha}$ values calculated by DFT for different exciton-phonon scattering processes that emits a phonon. These scattering processes are non-resonant with small $|S_{\nu\mu}^{\alpha}|$ values; the listed $|S_{\nu\mu}^{\alpha}|$ values are multiplied by a factor of 100. The approximate momentum of the emitted phonon is shown in the second column, where K, Q, M refer respectively to one of the K, Q, M (and symmetry-related) points. At each of these points, there are nine pairs of phonon modes, labeled by 1 – 9 from low to high energy; each pair is nearly degenerate. We have summed over the contributions of the two nearly degenerate phonon modes for the $S_{\nu\mu}^{\alpha}$ values listed here. The values for strong transitions are bolded. The number in the parenthese denote the energy of the emitted phonon (in unit of meV).

Below we will give some details of our calculation of $S_{\nu\mu}^{\alpha}$. The calculation involves the electron-phonon interaction, which we describe by the following operator: [20]

$$\hat{H}_{ep} = \sum_{nj\alpha\mathbf{q}} \sqrt{\frac{\hbar}{2NM_j\Omega_{\alpha\mathbf{q}}}} \nabla_j V(\mathbf{r}-\mathbf{R}_n-\mathbf{R}_j) \cdot \boldsymbol{\epsilon}_j^{\alpha} (a_{\alpha,-\mathbf{q}}^{+} + a_{\alpha\mathbf{q}}) e^{i\mathbf{q}\cdot\mathbf{R}_n} . \quad (8)$$

Here $N$ denotes the number of unit cells in the sample; $\mathbf{R}_n$ denotes the position of a unit cell; $\mathbf{R}_j$ and $M_j$ denote the position and mass of an atom at site $j$ in a unit cell; $V(\mathbf{r}-\mathbf{R}_n-\mathbf{R}_j)$ denotes the electron-ion interaction potential (screened by valence electrons) centered at $\mathbf{R}_n + \mathbf{R}_j$; $a_{\alpha-\mathbf{q}}^{+}$ ($a_{\alpha\mathbf{q}}$) creates (annihilates) a phonon of mode α and wave-vector **-q** (**q**); $\boldsymbol{\epsilon}_j^{\alpha}$ is the corresponding phonon polarization vector (*i.e.* the displacement unit vector of the atom at site *j*).

By using Eq. (8), we obtain

$$\langle X_\nu; \Omega_\alpha|\hat{H}_{ep}|X_\mu\rangle = O_{\nu\mu}\sum_j \sqrt{\frac{\hbar}{2NM_j\Omega_\sigma}}\,\Xi^j_{\mathbf{q}} \cdot \boldsymbol{\epsilon}^\alpha_j(\mathbf{q}). \quad (9)$$

Here $O_{\nu\mu} = \sum_{\mathbf{k}} \varphi^*_\nu(\mathbf{k})\,\varphi_\mu(\mathbf{k})$ denotes the overlap between the initial and mediating exciton state with respective k-space envelope functions $\varphi_\mu(\mathbf{k})$ and $\varphi_\nu(\mathbf{k})$, which satisfy the normalization condition $\sum_{\mathbf{k}} |\varphi_{\mu,\nu}(\mathbf{k})|^2 = 1$. Using a trial wave function of the form $e^{-\zeta r}$ to describe the exciton real-space envelope function, we obtain $O_{\nu\mu} = 4\zeta_\nu\zeta_\mu/\left(\zeta_\nu+\zeta_\mu\right)^2$, where $\zeta_\nu$ and $\zeta_\mu$ are the exponents of the corresponding ν and μ excitons. Since the exciton binding energy is proportional to $\zeta$ in a hydrogenic model, we have $O_{\nu\mu} = 4r_{\nu\mu}/\left(1 + r_{\nu\mu}\right)^2$, where $r_{\nu\mu} = E_{b,\mu}/E_{b,\nu}$ is the ratio of binding energies between the μ and ν excitons. Based on this formula, we estimate the $O_{\nu\mu}$ values for various exciton transitions (Table S4).

$$\Xi^j_{\mathbf{q}} = \sum_n \langle c_{1,\mathbf{Q}+\mathbf{q}}|\nabla_j V(\mathbf{r} - \mathbf{R}_n - \mathbf{R}_j)e^{i\mathbf{q}\cdot\mathbf{R}_n}|c_{1,\mathbf{Q}}\rangle = \sum_{\mathbf{G}} \mathbf{U}^j_{\mathbf{q}}(\mathbf{G})\langle u_{c1,\mathbf{Q}+\mathbf{q}}|e^{i(\mathbf{q}+\mathbf{G})\cdot\mathbf{r}}|u_{c1,\mathbf{Q}}\rangle \quad (10)$$

is the deformation vector induced by an atomic displacement at site $\mathbf{R}_j$ in each unit cell for the case of electron scattering ($c_1$ is replaced by $v_1$ for hole scattering).

Here

$$\mathbf{U}^j_{\mathbf{q}}(\mathbf{G}) = \frac{1}{A_c L_c}\sum_{\mathbf{G}} i(\mathbf{q} + \mathbf{G})\,\tilde{V}(\mathbf{q} + \mathbf{G})e^{-i\mathbf{G}\cdot\mathbf{R}_j}. \quad (11)$$

$\mathbf{G}$ denotes the reciprocal lattice vectors of the crystal; $\tilde{V}_j(\mathbf{q})$ is the Fourier transform of the crystal potential $V(\mathbf{r})$; $|u_{c1,\mathbf{Q}}\rangle$ denotes the periodic part of the Bloch state at $\mathbf{Q}$. In our semi-emipircal calculation, we approximate $\tilde{V}_j(\mathbf{q})$ as

$$\tilde{V}_j(\mathbf{q}) = -\frac{4\pi Z e^2}{q^2}\left[e^{-\left(sqR^j_{MT}\right)^2/4} - e^{-\left(s'qR^j_{MT}\right)^2/4}\right]. \quad (12)$$

Here Z denotes the valency number (6 for both W and Se); $R^j_{MT}$ denotes the muffintin radius of the atom $j$; and $s$ denotes an empirical parameter. The first term in Eq. (12) denotes the Coulomb potentential due to the ion with a gaussian charge distribution of radius $sR^j_{MT}$; the second term denotes the screening contribtion of valence electrons described by the same form but with radius $s'R^j_{MT}$. We determine the empirical parameters $s$ and $s'$ by fitting the deformation potential constant $D_0 = \sum_j \Xi^j_{\mathbf{0}} \cdot \boldsymbol{\epsilon}_j$ for electrons coupled to the Γ-point optical phonons. Prior DFT calculations have obtained $D_0 = 2.2\, eV/\text{Å}$ ($3.1\, eV/\text{Å}$) for the electrons in the top valance band at the Γ (K) point of monolayer $WSe_2$ [21]. By fitting these prior results, we obtain $s = 0.41$ and $s' = 1.1$.

We have used the LASTO codes to compute the integrals in Eq. (10) to obtain the deformation vector $\Xi_{\mathbf{q}}$. Afterward, by using the phonon polarization vectors $\boldsymbol{\epsilon}_j$ obtained by our rigid-ion model, we can calculate the exciton-phonon scattering strength $S^\alpha_{\nu\mu}$ according to Eqs. (6) and (9). The caluclated $S^\alpha_{\nu\mu}$ values for the relevant transitions are listed in Table S3.

After we obtain $S^{\alpha}_{\nu\mu}$, we move on to calculate the one-phonon replica spectra by Eq. (7). The calculation involes the electron-light interaction strength $|\langle\omega|\hat{H}_{el}|X_{\sigma}\rangle|^2$ for the three exciton states ($X_{KK}$, $X_{QQ}$, $X_{\Gamma\Gamma}$), averaged over their doubly energy-degenerate states (as illustrated in Fig. 3). We assume that the interaction strength is proportional to the absorption strength of the corrsponding transitions measured by spectroscopic ellipsometry as reported in Ref. [22]. The ratios of the $X_{KK}$, $X_{QQ}$, $X_{\Gamma\Gamma}$ oscillator strengths are estimated to be 1:2:4. By using these ratios of electron-light interaction strength, combined with the $S^{\alpha}_{\nu\mu}$ values, we obtain the spectra of the one-phonon replica (Fig. S10).

In our calculated replica spectra, the lowest energy peaks are only ~33 meV below the primrary exciton peak due to the limited energy range of single phonon in bilayer $WSe_2$. This cannot account for the replica peaks that reach redshift energy >40 meV in our expeirment. Therefore, we need to consider higher-order processes to explain our data.

| Transition | $X_{Q\Gamma} \rightarrow X_{Q'\Gamma}$<br>$X_{QK} \rightarrow X_{Q'K}$ | $X_{QK} \rightarrow X_{Q\Gamma}$ | $X_{QK} \rightarrow X_{KK}$ | $X_{QK} \rightarrow X_{QQ}$ | $X_{Q\Gamma} \rightarrow X_{QQ}$ | $X_{Q\Gamma} \rightarrow X_{\Gamma\Gamma}$ |
|---|---|---|---|---|---|---|
| $\lvert O_{\nu\mu}\rvert$ | 1 | 0.89 | 0.99 | 0.83 | 0.99 | 0.83 |

**Table S4.** The absolute value of the overlap integral $O_{\nu\mu}$ for various transitions between two excition states (denoted by μ and ν).

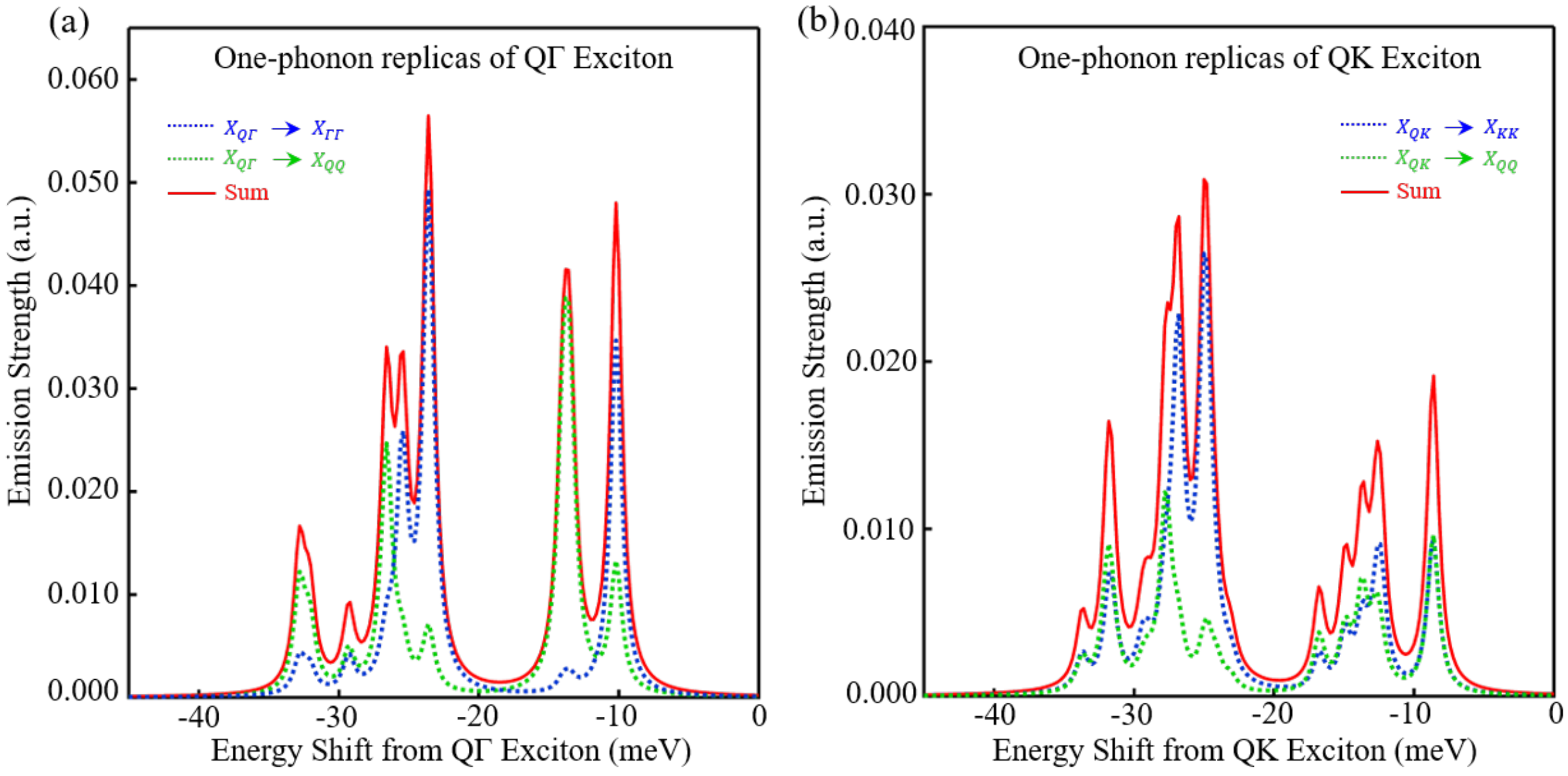

**Fig. S10.** Calculated one-phonon replica emission spectra for bilayer $WSe_2$. (a) One-phonon replica of $X_{Q\Gamma}$. (b) One-phonon replica of $X_{QK}$. We broaden the spectral lines by replacing the delta function in Eq. (7) with a Lorentizian function with half width of 0.5 meV.

Fig. S10(a) displays the calculated one-phonon replica spectrum of QΓ exciton. In our plot, we appropriately broaden the spectral lines by replacing the delta function in Eq. (7) with a Lorentizian function with half width of 0.5 meV. The spectrum is contributed by the $X_{Q\Gamma} \rightarrow X_{QQ}$ and $X_{Q\Gamma} \rightarrow X_{\Gamma\Gamma}$ transitions, whose constituent spectra are also shown for comparison. Four prominent peaks are found at ~10, ~14.0, 23.6, and 25.4 meV below the zero-phonon line. The two peaks at ~10 and ~14 meV are close to the broad $X_{Q\Gamma}^{1}$ replica at 13.0 meV in our experiment. The two peaks at 23.6 and 25.4 meV are close to the $X_{Q\Gamma}^{2}$ replica at 27.6 meV in experiment. Fig. S10(b) displays the one-phonon replica spectrum of the QK exciton, which is contributed by the $X_{QK} \rightarrow X_{QQ}$ and $X_{QK} \rightarrow X_{KK}$ transitions.

### 5.6 Calculation of two-phonon replica spectra

The emission strength of two-phonon-assisted intervalley exciton recomination is proportional to the transition rate ($P^{(2)}$) according to Fermi's golden rule and the perturbation theory:

$$P_{\mu}^{(2)}(\omega) = \frac{2\pi}{\hbar} \sum_{\mathbf{q},\alpha,\beta} \left| \sum_{\nu,\sigma} \frac{\langle \omega\,; \Omega_{\alpha\mathbf{q}}, \Omega_{\beta} | \hat{H}_{el} | X_{\sigma}; \Omega_{\alpha\mathbf{q}}, \Omega_{\beta} \rangle \langle X_{\sigma}; \Omega_{\alpha\mathbf{q}}, \Omega_{\beta} | \hat{H}_{ep} | X_{\nu}; \Omega_{\alpha\mathbf{q}} \rangle \langle X_{\nu}; \Omega_{\alpha\mathbf{q}} | \hat{H}_{ep} | X_{\mu} \rangle}{\left(E_{\nu} - E_{\sigma} - \hbar\Omega_{\beta} + i\gamma\right)\left(E_{\mu} - E_{\nu} - \hbar\Omega_{\alpha\mathbf{q}} + i\gamma\right)} \right|^{2} \cdot \delta\left(E_{\mu} - \hbar\Omega_{\alpha\mathbf{q}} - \hbar\Omega_{\beta} - \hbar\omega\right). \quad (13)$$

Here $\hat{H}_{el}$ ($\hat{H}_{ep}$) is the electron-light (electron-phonon) interaction Hamiltonian; ω denotes the frequency of emitted photon; $X_{\mu}$ with μ = QΓ or QK denotes the initial intervalley exciton state with energy $E_{\mu}$; $X_{\nu}$ ($X_{\sigma}$) denotes the first (second) mediating exciton state with energy $E_{\nu}$ ($E_{\sigma}$); $\Omega_{\alpha\mathbf{q}}$ and $\Omega_{\beta}$ denotes the frequencies of the two emitted phonons in modes α and β; γ = 2 meV is a phenomenological broadening energy due to the finite carrier lifetime.

We note that, in the two-phonon process, the wave vector **q** of the first emitted phonon ($\Omega_{\alpha\mathbf{q}}$) can vary; therefore, we need to sum over states corresponding to different **q**. However, once the wave vector of the first phonon is given, the wave vector of the second phonon ($\Omega_{\beta}$) is determined by the conservation of momentum in the two-phonon scattering process. Therefore Eq. (13) does not contain the sum over the wave vector of the second phonon. For the dominant two-phonon processes, the first phonon is coupled (nearly) resonantly to the exciton states, hence the scattering strength is sensitive to a small variation of the phonon wave vector **q**; but the second phonon is coupled non-resonantly to the exciton states, hence the scattering strength is insensitive to a small variation of the phonon wave vector.

In Eq. (13), the $\hat{H}_{ep}$ matrix elements in the numerator are insensitive to **q**, but the inverse energy factor in the denominator is sensitive to **q** for the (nearly) resonant process. Therefore, we can simplify the expression by approximating the numerator at a fixed **q** corresponding to the K, Q or Γ point, while suming over the inverse energy factor over **q**. By using this approach and summing over energy-degenerate excitonic transitions, Eq. (13) can be simplified as

$$P_{\mu}^{(2)}(\omega) = \frac{2\pi}{N\hbar}\sum_{\alpha,\beta}\sum_{\nu,\sigma} f_{\sigma\nu\mu}^{\alpha}\left|\langle\omega|\widehat{H}_{el}|X_{\sigma}\rangle\right|^{2} S_{\sigma\nu}^{\beta} S_{\nu\mu}^{\alpha}\,\delta\!\left(E_{\mu} - \hbar\Omega_{\alpha} - \hbar\Omega_{\beta} - \hbar\omega\right). \quad (14)$$

Here $S_{\nu\mu}^{\alpha}$ and $S_{\sigma\nu}^{\beta}$ are the exciton-phonon scattering strengths defined by Eq. (6), evalulated at a fixed **q** that corresponds to the K, Q, or Γ point. The integral over **q** is included in the factor $f^{\alpha}$, which is given as:

$$f_{\sigma\nu\mu}^{\alpha} \approx \frac{A_c g_{\sigma\nu\mu}^{2}}{(2\pi)^2}\int_0^{\infty} d^2q \left|\frac{\Delta_{\alpha} + i\gamma}{\Delta_{\alpha} - \frac{\hbar^2 q^2}{2m_{\nu}} + i\gamma}\right|^2 = \frac{A_c g_{\sigma\nu\mu}^{2}|\Delta_{\alpha} + i\gamma|^2 m_{\nu}}{2\pi\hbar^2\gamma}\left[\frac{\pi}{2} + \tan^{-1}\left(\frac{\Delta_{\alpha}}{\gamma}\right)\right] \quad (15)$$

Here $A_c = 32.75$ (in atomic unit) is the area of the primitive cell; $m_{\nu}$ is the effective mass of the exction state ν (in unit of free eletron mass); $\Delta_{\alpha} = E_{\mu} - E_{\nu} - \hbar\Omega_{\alpha}$ ; $g_{\sigma\nu\mu}$ is a degeneracy factor, which is the number of degenrate transitions emitting the same phonons.

In order to obtain $P_{\mu}^{(2)}$ in Eq. (14), we need to calculate $S$ for different transitions. Below we will consider the dominant two-phonon transition processes.

For both QΓ and QK initial exciton state, the second mediating state should be a momentum-direct exciton state that couples to light. The dominating direct exciton state should be at either K, Q, or Γ valley (or region) (denoted as $X_{KK}$, $X_{QQ}$, $X_{\Gamma\Gamma}$, respectively), because they are all strongly coupled to light.

For the QΓ initial exciton, there are four dominating groups of two-phonon pathsways to reach such a direct exciton state. These pathways dominante because they all invovle one nearly resonant mediating state. Below we will describe each group of pathways.

(1) $X_{Q\Gamma} \rightarrow X_{Q'\Gamma} \rightarrow X_{Q'Q'}$

In the first group of pathways, $X_{Q\Gamma}$ first transits to $X_{Q'\Gamma}$ through electron scattering from Q to Q' valley (this is a nearly resonant transition because Q and Q' valleys are energy degenerate); a phonon with momentum close to **Q – Q'** is emitted in this process (this phonon can be near the Q, K or M point, or their symmetry-related points). Afterward, $X_{Q'\Gamma}$ transits to $X_{Q'Q'}$ through hole scattering from Γ valley to Q' region; a second phonon with momentum close to **–Q'** is emitted in this process. Here Q' denotes any of the six inequivalent Q valleys; therefore this group of pathways includes six different paths via six different Q' valleys. We note that when Q' is Q, the $X_{Q\Gamma} \rightarrow X_{Q\Gamma}$ transition is an intravalley transition that emits a phonon near the Γ point.

(2) $X_{Q\Gamma} \rightarrow X_{Q'\Gamma} \rightarrow X_{\Gamma\Gamma}$

In the second group of pathways, the $X_{Q\Gamma} \rightarrow X_{Q'\Gamma}$ transition is the same as in the first group. In the second transition, $X_{Q'\Gamma}$ transits to $X_{\Gamma\Gamma}$ through electron scattering from Q' valley to Γ region in the conduction band; a second phonon with momentum close to **Q'** is emitted in this process. This group includes six pathways via six different Q' valleys.

(3) $X_{Q\Gamma} \rightarrow X_{QK} \rightarrow X_{KK}$

In the third group of pathways, $X_{Q\Gamma}$ first transits to $X_{QK}$ through hole scattering from Γ to K

valley; a phonon with momentum close to –**K** is emitted in this process (this is a resonant or nearly resonant transition because of the samll $X_{Q\Gamma} - X_{QK}$ energy separation). Afterward, $X_{QK}$ transits to $X_{KK}$ through electron scattering from Q to K valley; a second phonon with momentum close to **Q** – **K** ≈ –**Q** is emitted. We note that there are two inquivalent K valleys (*i.e.* K and K' valleys); so, this group includes two pathways via the K or K' valley.

(4) $X_{Q\Gamma} \rightarrow X_{Q\mathrm{K}} \rightarrow X_{Q\mathrm{Q}}$

In the fourth group of pathways, the $X_{Q\Gamma} \rightarrow X_{Q\mathrm{K}}$ transition is the same as in the third group. In the second transition, $X_{Q\mathrm{K}}$ transits to $\mathrm{X_{QQ}}$ through hole scattering from K valley to Q region in the valence band; a second phonon with momentum close to **K** – **Q** ≈ **Q** is emitted in this process. This group includes two pathways via the K or K' valley.

By including all of the four types of pathways above, there are totally 16 (nearly) resonant pathways in the two-phonon process of $X_{Q\Gamma}$.

For the QK initial exciton, there are also four dominating groups of two-phonon pathways to reach a momentum-direct exciton state. These pathways dominante because they all invovle one (nearly) resonant mediating state. Below we will describe each group of pathways.

(1) $X_{Q\mathrm{K}} \rightarrow X_{Q'\mathrm{K}} \rightarrow \mathrm{X_{KK}}$

In the first group of pathways, $X_{Q\mathrm{K}}$ first transits to $X_{Q'\mathrm{K}}$ through electron scattering from Q to Q' valley; a phonon with momentum close to **Q** – **Q'** is emitted in this process (this is a nearly resonant transition because Q and Q' valleys are energy degenerate). Afterward, $X_{Q'\mathrm{K}}$ transits to $\mathrm{X_{KK}}$ through electron scattering from Q' to K valley; a second phonon with momentum close to **Q'** – **K** is emitted in this process. This group includes six different paths via six different Q' valleys. When Q' is Q, the $X_{Q\mathrm{K}} \rightarrow X_{Q\mathrm{K}}$ transition is an intravalley transition that emits a phonon near the Γ point.

(2) $X_{Q\mathrm{K}} \rightarrow X_{Q'\mathrm{K}} \rightarrow \mathrm{X}_{Q'Q'}$

In the second group of pathways, the $X_{Q\mathrm{K}} \rightarrow X_{Q'\mathrm{K}}$ transition is the same as in the first group. In the second transition, $X_{Q'K}$ transits to $\mathrm{X}_{Q'Q'}$ through hole scattering from K valley to Q' region in the valence band; a second phonon with momentum close to **K** – **Q'** is emitted in this process. This group includes six pathways via six different Q' valleys.

(3) $X_{Q\mathrm{K}} \rightarrow X_{Q\Gamma} \rightarrow X_{QQ}$

In the third group of pathways, $X_{Q\mathrm{K}}$ first transits to $X_{Q\Gamma}$ through hole scattering from K to Γ valley; a phonon with momentum close to **K** is emitted in this process (this is a resonant or nearly resonant transition because the $X_{Q\Gamma} - X_{QK}$ energy separation is close to the energy of a phonon). Afterward, $X_{Q\Gamma}$ transits to $X_{QQ}$ through hole scattering from Γ to Q valley; a second phonon with momentum –**Q** is emitted. There is only one pathway in this group.

(4) $X_{QK} \to X_{Q\Gamma} \to X_{\Gamma\Gamma}$

In the fourth group of pathways, the $X_{QK} \to X_{Q\Gamma}$ transition is the same as in the third group. In the second transition, $X_{Q\Gamma}$ transits to $X_{\Gamma\Gamma}$ through electron scattering from Q valley to Γ region in the conduction band; a second phonon with momentum close to **Q** is emitted. This group includes only one pathway.

By including all of the four types of pathways above, there are totally 14 dominating pathways in the two-phonon process of $X_{QK}$.

We note that for all of the above pathways for QΓ and QK initial excitons, the emitted phonons are all near the Γ, Q, M, and K points (and their symmetry-related points). These faciliate our assignment of the phonon replicas.

| **Exction-phonon transitions** | **Phonon momentum** | **Phonon branches from low to high energy** | | | | | | | | |
|---|---|---|---|---|---|---|---|---|---|---|
| | | **1** | **2** | **3** | **4** | **5** | **6** | **7** | **8** | **9** |
| $X_{Q_1\Gamma} \to X_{Q_1\Gamma}$ | Γ | 0<br>(0) | 0<br>(0) | 0<br>(0) | 1.933<br>(21.4) | **5.677**<br>**(21.5)** | 0.284<br>(29.9) | 0.079<br>(29.9) | **12.97**<br>**(30.2)** | **2.550**<br>**(37.1)** |
| $X_{Q_1K} \to X_{Q_1K}$ | Γ | 0<br>(0) | 0<br>(0) | 0<br>(0) | 1.931<br>(21.4) | **5.670**<br>**(21.5)** | 0.284<br>(29.9) | 0.078<br>(29.9) | **18.18**<br>**(30.2)** | 0.230<br>(37.1) |
| $X_{Q_1\Gamma} \to X_{Q'_1\Gamma}$<br>$X_{Q_1K} \to X_{Q'_1K}$ | K | **12.22**<br>**(13.1)** | 0.597<br>(15.3) | **6.637**<br>**(17.6)** | 0.156<br>(23.8) | 0.304<br>(26.0) | **2.936**<br>(26.6) | 0.852<br>(29.7) | 0.245<br>(30.9) | **1.586**<br>**(32.4)** |
| $X_{Q_1\Gamma} \to X_{Q'_2\Gamma}$<br>$X_{Q_1\Gamma} \to X_{Q'_3\Gamma}$<br>$X_{Q_1K} \to X_{Q'_2K}$<br>$X_{Q_1K} \to X_{Q'_3K}$ | Q | **1.168**<br>**(10.1)** | **3.310**<br>**(13.5)** | 0.628<br>(13.9) | **2.298**<br>**(23.6)** | 0.204<br>(25.4) | 0.603<br>(26.6) | 0.134<br>(29.3) | 0.321<br>(32.1) | 0.086<br>(32.8) |
| $X_{Q_1\Gamma} \to X_{Q_2\Gamma}$<br>$X_{Q_1\Gamma} \to X_{Q_3\Gamma}$<br>$X_{Q_1K} \to X_{Q_2K}$<br>$X_{Q_1K} \to X_{Q_3K}$ | M | **9.813**<br>**(13.7)** | **2.762**<br>**(14.6)** | 0.057<br>(16.6) | 0.007<br>(24.0) | 1.694<br>(24.9) | **5.293**<br>**(27.7)** | 0.025<br>(29.0) | 0.830<br>(31.6) | 0.345<br>(31.8) |
| $X_{QK} \to X_{Q\Gamma}$<br>for any Q valley | K | **20.55**<br>**(13.0)** | **124.5**<br>**(15.4)** | **106.4**<br>**(17.7)** | 2.497<br>(23.7) | 4.683<br>(26.1) | **19.66**<br>**(26.6)** | 5.135<br>(29.7) | 6.306<br>(30.8) | 0.477<br>(32.4) |
| $X_{Q\Gamma} \to X_{QK}$<br>for any Q valley | K | 0.585<br>(13.0) | **1.109**<br>**(15.4)** | 0.338<br>(17.7) | 0.056<br>(23.7) | 0.178<br>(26.1) | 0.811<br>(26.6) | 0.333<br>(29.7) | 0.463<br>(30.8) | 0.041<br>(32.4) |

**Table S5.** The $S^{\alpha}_{\nu\mu}$ values calculated by DFT for different exciton-phonon scattering processes that emit the first phonon in the two-phonon replica. These scattering processes are all (nearly) resonant. The approximate momentum of the emitted phonon is listed in the second column, where K denotes the K or K' point, Q denotes any of the six Q points, and M denotes any of the three M points ($M_1$, $M_2$, $M_3$) in the Brillouin zone. At each of these points, there are nine pairs of phonon modes, labeled by 1 – 9 from low to high energy; each pair is nearly degenerate. We have summed over the contributions from the two nearly degenearte phonon modes for the $S^{\alpha}_{\mu\nu}$ values listed here. The values for strong transitions are bolded. The number in the parenthese denote the energy of the emitted phonon (in unit of meV).

For all of the two-phonon pathways listed above, they consist of two transitions – the first transition emits the first phonon through a (nearly) resonant scattering process, and the second transition emits the second phonon through a non-resonant scattering process. We have calculated the exciton-phonon scattering strength for both transitions. The scattering strength of the first (resonant or nearly resonant) process is listed in Table S5 for all the relevant pathways. The second (non-resonant) process is the same as those in the single-phonon replicas; the scattering strength has been listed in Table S3. By combining Tables S3 and S5, we can obtain the strength of the two-phonon processes as the product of the strengths of its two constituent one-phonon processes.

After we obtain the two-phonon scattering strength $S_{\sigma\nu}^{\beta} S_{\nu\mu}^{\alpha}$, we calculate the two-phonon replica spectra by using Eq. (14) and the 1:2:4 ratio of $\left|\langle\omega|\hat{H}_{el}|X_\sigma\rangle\right|^2$ for $X_{KK}$, $X_{QQ}$, $X_{\Gamma\Gamma}$, similar to our calculation of single-phonon replica. The calculated two-phonon replica spectra for $X_{Q\Gamma}$ and $X_{QK}$ are displayed in Fig. S11. In Fig. S11, we broaden the spectral lines by replacing the delta function in Eq. (14) with a Lorentizian function with half width of 0.5 meV.

For the two-phonon replicas of QΓ exciton, our calculated spectrum shows multiple peaks, which can somewhat account for $X_{Q\Gamma}^{1-5}$ observed in our experiment [see the labels in Fig. S11(a)]. In our theoretical results, the two-phonon replicas for $X_{Q\Gamma}$ are mainly contributed by the $X_{Q\Gamma} \rightarrow X_{Q'\Gamma} \rightarrow X_{Q'Q'}$ and $X_{Q\Gamma} \rightarrow X_{Q'\Gamma} \rightarrow X_{\Gamma\Gamma}$ paths, while the other two types of paths ($X_{Q\Gamma} \rightarrow X_{QK} \rightarrow X_{KK}$; $X_{Q\Gamma} \rightarrow X_{QK} \rightarrow X_{QQ}$) only contribute weakly. Notably, $X_{Q\Gamma}^5$ is contributed significantly by two-phonon paths involving the $X_{Q\Gamma} \rightarrow X_{Q\Gamma}$ intravalley transition, which emits an optical phonon near the Γ point.

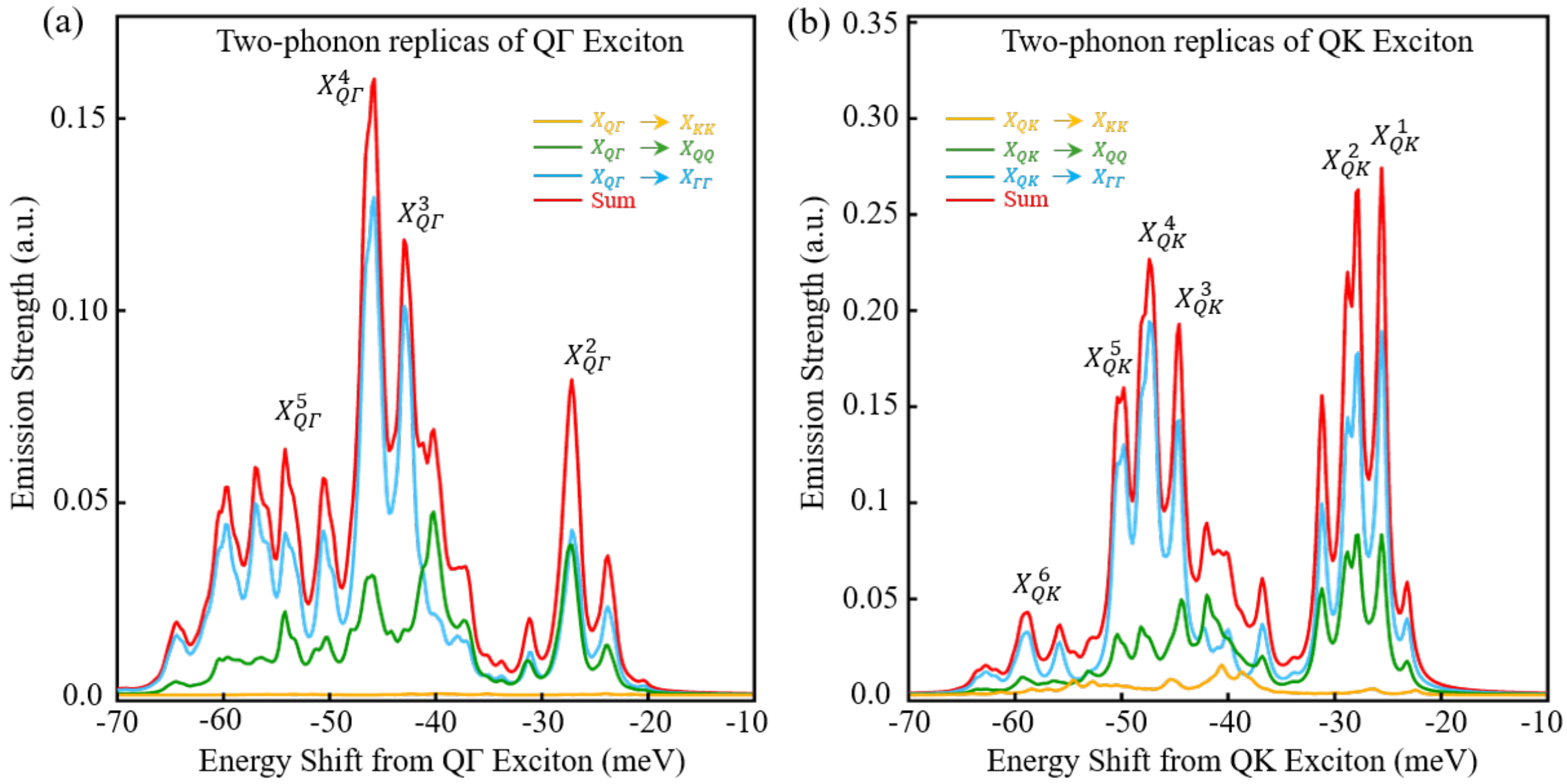

**Fig. S11.** Calculated emission spectra of two-phonon replicas for (a) QΓ exciton and (b) QK exciton in bilayer $WSe_2$. The figure displays the spectra associated with different types of two-phonon pathways as well as the total spectra as their sum.

For the two-phonon replicas of QK exciton, our calculated spectrum shows multiple peaks, which can somewhat account for $X_{QK}^{1-6}$ observed in our experiment [see the labels in Fig. S11(b)]. In our theoretical results, the two-phonon replicas for $X_{QK}$ are mainly contributed by the $X_{QK} \rightarrow X_{Q\Gamma} \rightarrow X_{QQ}$ and $X_{QK} \rightarrow X_{Q\Gamma} \rightarrow X_{\Gamma\Gamma}$ pathways. Notably, the $X_{QK} \rightarrow X_{Q\Gamma}$ transition with acoustic phonon emission is strongly resonant when the emitted phonon energy is close to the 18-meV difference between $X_{QK}$ and $X_{Q\Gamma}$ (Table S5). Two-phonon processes involving such strongly resonant transitions contribute dominantly to $X_{QK}^{1-5}$ in our result. $X_{QK}^{6}$ is contributed significantly by two-phonon pathways involving the $X_{QK} \rightarrow X_{QK}$ intravalley transition, which emits an optical phonon near the Γ point.

Fig. S12 displays the calculated total one-phonon and two-phonon replica spectra for both $X_{Q\Gamma}$ and $X_{QK}$. The two-phonon replicas are overall stronger than the one-phonon replicas. There are two factors for this behavior. First, the additional transition in the two-phonon process is a (nearly) resonant process, which has a transition rate close to 100%. Second, there are many more pathways in the two-phonon processes (≥14 paths) than in the one-phonon processes (only two paths); thus, the total transition rate is proprotionally larger in the two-phonon processes.

In Figs. S10-12, we only mildly broaden the spectra by replacing the delta function in Eqs. (7) and (14) with a Lorentizian function of half width 0.5 meV. The theorectical spectra hence display sharp lines. In our experiment, however, the replica peaks are relatively broad. To compare with the experimental results, we plot the theorectial one-phonon, two-phonon, and total replica spectra by using a Lorentizian broadening function of half width 2 meV. The results, displayed in Fig. 4 of the main paper, agree decently with our experimental spectra.

Our theoretical simulation enables us to give detailed assignments of the observed phonon replicas. In Table S6, we list the relevant exciton-phonon scattering processes and phonon modes for different replica peaks for both QΓ and QK excitons. Further research is merited to resolve replicas associated with individual scattering process and further understand the rich exciton-phonon coupling physics in bilayer $WSe_2$.

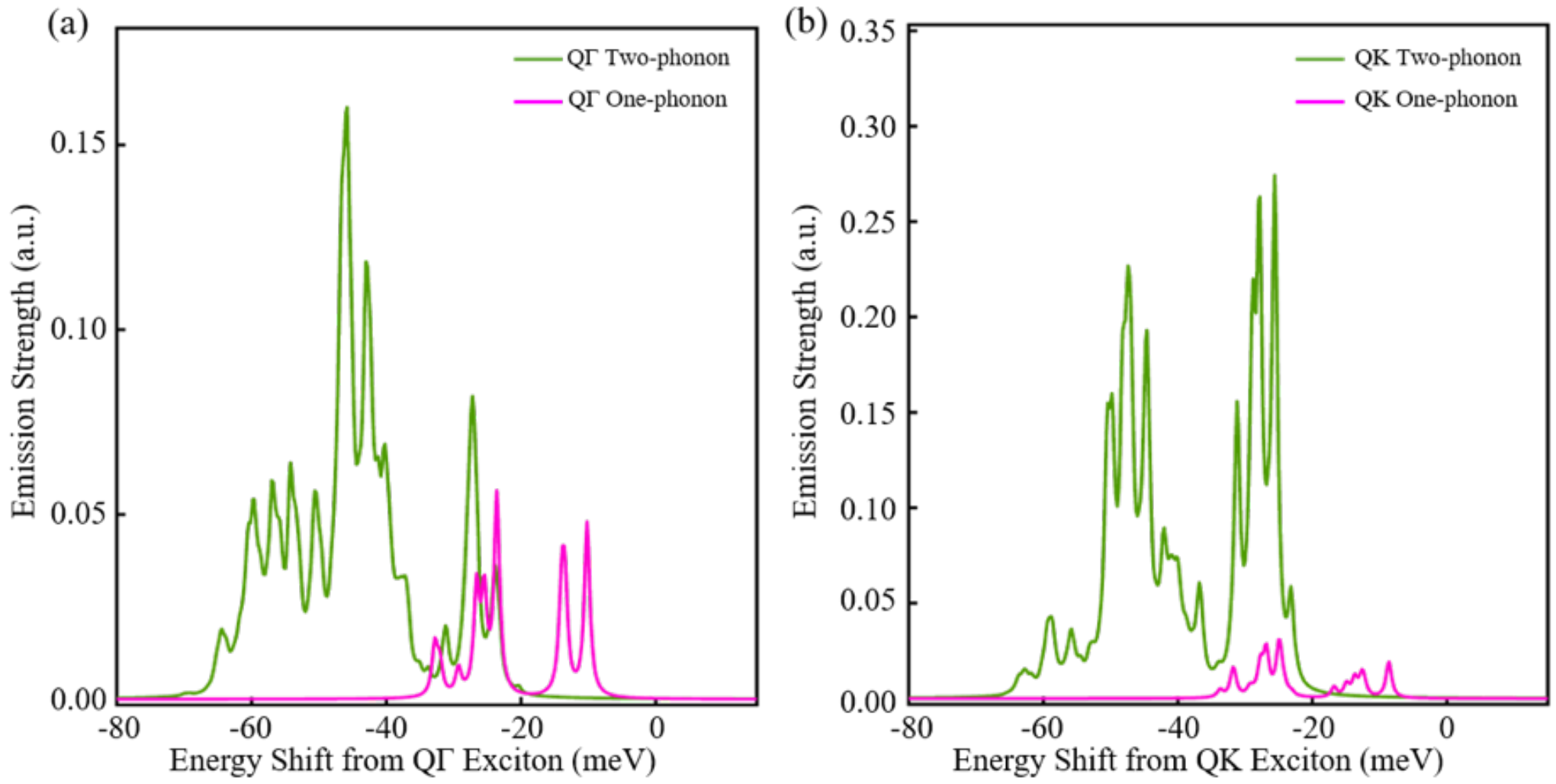

**Fig. S12.** Comparison of relative emission strength of one-phonon replica and two-phonon replica for (a) QΓ exciton and (b) QK exciton in bilayer $WSe_2$.

| Phonon replicas | Redshift (Expt.) (meV) | Redshift (Theory) (meV) | First transition | | Second transition | |
|---|---|---|---|---|---|---|
| | | | Transition | Phonon mode (energy in meV) | Transition | Phonon mode (energy in meV) |
| $X_{Q\Gamma}^1$ | 13.0 ± 2.0 | 13.5 | $X_{Q\Gamma} \rightarrow X_{QQ}$ | Q2 (13.5)<br>Q3 (14.0) | – | – |
| $X_{Q\Gamma}^2$ | 27.6 ± 1.0 | 24.0 | $X_{Q\Gamma} \rightarrow X_{QK}$ | K2 (15.4) | $X_{QK} \rightarrow X_{QQ}$ | Q1 (8.57) |
| | | 26.3 | $X_{Q\Gamma} \rightarrow X_{QK}$ | K3 (17.7) | $X_{QK} \rightarrow X_{QQ}$ | Q1 (8.57) |
| | | 27.2 | $X_{Q_1\Gamma} \rightarrow X_{Q_2\Gamma}$<br>$X_{Q_1\Gamma} \rightarrow X_{Q_3\Gamma}$ | M1 (13.8) | $X_{Q_2\Gamma} \rightarrow X_{Q_2Q_2}$<br>$X_{Q_3\Gamma} \rightarrow X_{Q_3Q_3}$<br>$X_{Q_2\Gamma} \rightarrow X_{\Gamma\Gamma}$<br>$X_{Q_3\Gamma} \rightarrow X_{\Gamma\Gamma}$ | Q2 (13.5) |
| $X_{Q\Gamma}^3$ | 41.7 ± 2.0 | 40.3 | $X_{Q_1\Gamma} \rightarrow X_{Q_2\Gamma}$<br>$X_{Q_1\Gamma} \rightarrow X_{Q_3\Gamma}$ | M1 (13.8) | $X_{Q_2\Gamma} \rightarrow X_{Q_2Q_2}$<br>$X_{Q_3\Gamma} \rightarrow X_{Q_3Q_3}$<br>$X_{Q_2\Gamma} \rightarrow X_{\Gamma\Gamma}$<br>$X_{Q_3\Gamma} \rightarrow X_{\Gamma\Gamma}$ | Q5 (26.6) |
| | | 40.4 | $X_{Q\Gamma} \rightarrow X_{QK}$ | K2 (15.4) | $X_{QK} \rightarrow X_{QQ}$ | Q5 (25.0) |
| | | 42.2 | $X_{Q\Gamma} \rightarrow X_{QK}$ | K2 (15.4) | $X_{QK} \rightarrow X_{QQ}$ | Q6 (26.8) |
| $X_{Q\Gamma}^4$ | 45.7 ± 1.2 | 44.5 | $X_{Q\Gamma} \rightarrow X_{QK}$ | K3 (17.7) | $X_{QK} \rightarrow X_{QQ}$ | Q6 (26.8) |
| | | 45.9 | $X_{Q_1\Gamma} \rightarrow X_{Q_1'\Gamma}$ | K1 (13.1) | $X_{Q_1'\Gamma} \rightarrow X_{Q_1'Q_1'}$ | Q9 (32.8) |
| | | 46.6 | $X_{Q_1\Gamma} \rightarrow X_{Q_2\Gamma}$<br>$X_{Q_1\Gamma} \rightarrow X_{Q_3\Gamma}$ | M1 (13.8) | $X_{Q_2\Gamma} \rightarrow X_{Q_2Q_2}$<br>$X_{Q_3\Gamma} \rightarrow X_{Q_3Q_3}$<br>$X_{Q_2\Gamma} \rightarrow X_{\Gamma\Gamma}$<br>$X_{Q_3\Gamma} \rightarrow X_{\Gamma\Gamma}$ | Q9 (32.8) |
| $X_{Q\Gamma}^5$ | 57.7 ± 1.4 | 56.9<br>59.6 | $X_{Q\Gamma} \rightarrow X_{Q\Gamma}$ | Γ8 (30.3) | $X_{Q\Gamma} \rightarrow X_{QQ}$ | Q6 (26.6)<br>Q7(29.3) |

| Phonon replicas | Redshift (Expt.) (meV) | Redshift (Theory) (meV) | First transition | | Second transition | |
|---|---|---|---|---|---|---|
| | | | Transition | Phonon mode (energy in meV) | Transition | Phonon mode (energy in meV) |
| $X_{QK}^1$ | 28.8 ± 1.8 | 25.6 | $X_{QK} \rightarrow X_{Q\Gamma}$ | K2 (15.4) | $X_{Q\Gamma} \rightarrow X_{\Gamma\Gamma}$ | Q1 (10.2) |
| | | 27.9 | $X_{QK} \rightarrow X_{Q\Gamma}$ | K3 (17.7) | $X_{Q\Gamma} \rightarrow X_{\Gamma\Gamma}$ | Q1 (10.2) |
| | | 28.9 | $X_{QK} \rightarrow X_{Q\Gamma}$ | K2 (15.4) | $X_{Q\Gamma} \rightarrow X_{QQ}$ | Q2 (13.5) |
| $X_{QK}^2$ | 32.4 ± 1.2 | 31.2 | $X_{QK} \rightarrow X_{Q\Gamma}$ | K3 (17.7) | $X_{Q\Gamma} \rightarrow X_{\Gamma\Gamma}$ | Q2 (13.5) |
| $X_{QK}^3$ | 41.9 ± 1.8 | 42.0 | $X_{QK} \rightarrow X_{Q\Gamma}$ | K2 (15.4) | $X_{Q\Gamma} \rightarrow X_{QQ}$ | Q6(26.6) |
| $X_{QK}^4$ | 46.2 ± 1.8 | 47.4 | $X_{QK} \rightarrow X_{Q\Gamma}$ | K2 (15.4) | $X_{Q\Gamma} \rightarrow X_{\Gamma\Gamma}$ | Q8 (32.0) |
| | | 44.3 | $X_{QK} \rightarrow X_{Q\Gamma}$ | K3 (17.7) | $X_{Q\Gamma} \rightarrow X_{\Gamma\Gamma}$ | Q6 (26.6) |
| $X_{QK}^5$ | 49.1 ± 1.6 | 48.2 | $X_{QK} \rightarrow X_{Q\Gamma}$ | K2 (15.4) | $X_{Q\Gamma} \rightarrow X_{\Gamma\Gamma}$ | Q9 (32.8) |
| | | 50.5 | $X_{QK} \rightarrow X_{Q\Gamma}$ | K3 (17.7) | $X_{Q\Gamma} \rightarrow X_{\Gamma\Gamma}$ | Q9 (32.8) |
| $X_{QK}^6$ | 58.3 ± 2.0 | 59.4 | $X_{QK} \rightarrow X_{QK}$<br>$X_{QK} \rightarrow X_{Q\Gamma}$ | Γ8 (30.3)<br>K6 (26.6) | $X_{QK} \rightarrow X_{KK}$<br>$X_{Q\Gamma} \rightarrow X_{QQ}$ | M7 (29.0)<br>Q9 (32.8) |

**Table S6.** Dominant transitions and phonon modes involved in different replica features for QΓ and QK excitons in bilayer $WSe_2$. All the listed energies are in unit of meV. The errors of the experimental energies represent a confidence interval of about 95%.

**Supplemental References:**